\documentclass[10pt,twocolumn,letterpaper]{article}

\usepackage[pagenumbers]{cvpr} 
\usepackage{xcolor}
\usepackage{colortbl}
\usepackage{minted}
\usepackage{multirow}
\usepackage{booktabs}
\usepackage{caption}
\usepackage{subcaption}
\usepackage{wrapfig}

\definecolor{cvprblue}{rgb}{0.21,0.49,0.74}
\usepackage[pagebackref,breaklinks,colorlinks,allcolors=cvprblue]{hyperref}

\definecolor{diffred}{RGB}{255,200,200} 
\definecolor{diffgreen}{RGB}{200,255,200} 
\definecolor{diffyellow}{RGB}{255,255,200} 
\newcommand{\xmark}{\cellcolor{diffred}\textcolor{red}{$\times$}}
\newcommand{\cmark}{\cellcolor{diffgreen}\textcolor{green}{$\checkmark$}}

\def\paperID{*****} 
\def\confName{CVPR}
\def\confYear{2027}

\title{DiffRayve: Differentiable Ray-Wave Method for Polarized and Unpolarized Diffractive-Refractive Optical Systems} 

\author{Samuel Audia\\
University of Maryland, College Park\\
Department of Computer Science\\
{\tt\small sjaudia@umd.edu}
\and
Shrey Patel\\
University of Maryland, College Park \\
Department of Computer Science\\
{\tt\small pshrey@umd.edu}
\and
Dinesh Manocha\\
University of Maryland, College Park \\
Department of Computer Science\\
Department of Electrical and Computer Engineering \\
{\tt\small dmanocha@umd.edu}
\and
Matthias Zwicker\\
University of Maryland, College Park \\
Department of Computer Science\\
{\tt\small zwicker@umd.edu}
}

\begin{document}
\maketitle
\begin{abstract} 

Current deep optics simulations struggle to balance high fidelity with computational efficiency, particularly for complex polarized and diffractive systems. To address this, we introduce a fully differentiable Shooting and Bouncing Ray (SBR) algorithm that simultaneously models geometric and physical optics fields. Unlike standard wave optics methods, our approach enables accurate, gradient-based optimization of polarized compound refractive-diffractive systems without the computational cost of full-wave simulations. We validate our method against analytical Fraunhofer diffraction patterns and state-of-the-art Fourier optics and wave optics methods. We showcase the practical utility of our differentiable engine through one application: designing a wide field-of-view achromatic lens with multiple Diffractive Optical Elements (DOEs).

\end{abstract}

\section{Introduction}
\begin{figure*}
  \centering
  \includegraphics[width=1.0\linewidth]{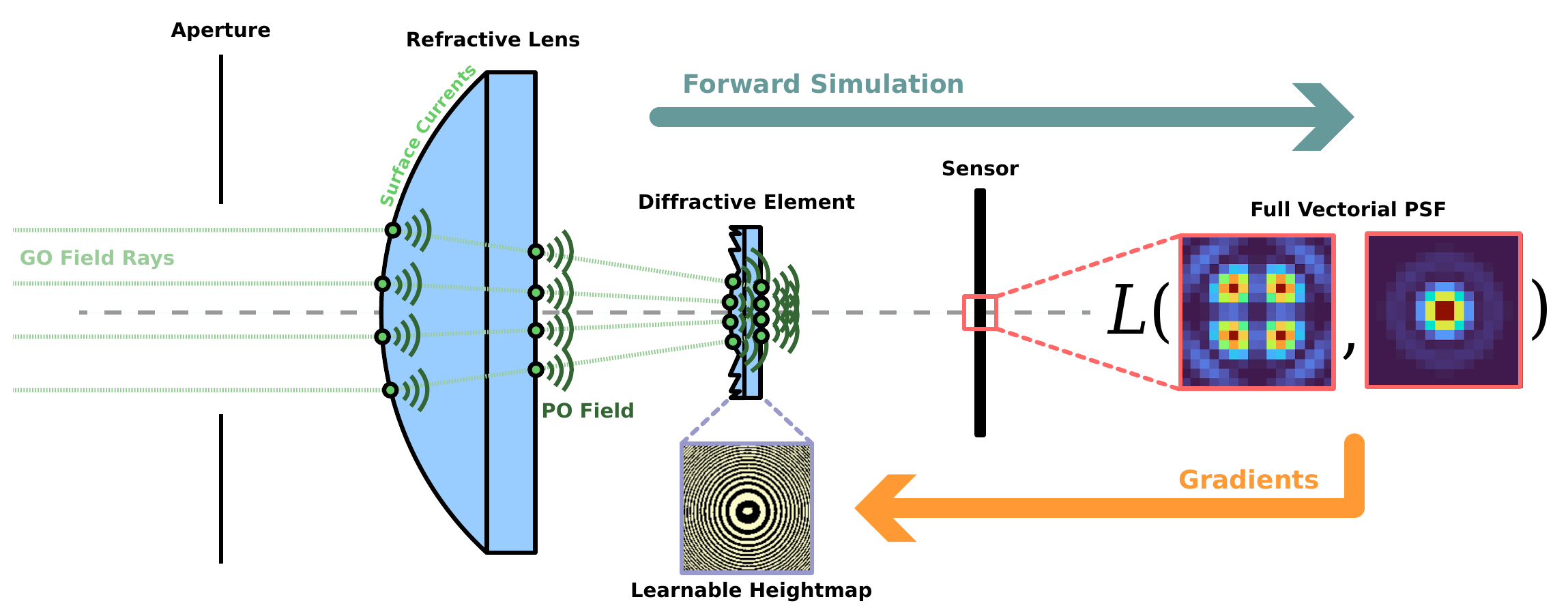}
  \caption{\textbf{DiffRayve Illustrative Example.} We develop a differentiable shooting and bouncing ray algorithm (SBR) that seamlessly captures GO and PO fields to model the ray and wave nature of light. Our method is able to optimize refractive and diffractive elements anywhere in the system and can handle polarized and unpolarized light. GO rays
    are traced through the lens system and at every intersection an
    equivalent current is calculated.
    Though not depicted, we sample both reflected and refracted
    rays using Monte Carlo integration. The surface currents are
    propagated to the
    sensor from all lens surfaces using
    a PO-like integral to calculate a full 3D, complex vector point spread
    function. The height map of
    diffractive elements can be optimized using
    the differentiable simulation to meet the lens designer's
    objective. Our approach allows for any combination of refractive
  and diffractive elements and is not restricted to the system shown above.\label{fig:overview}}
\end{figure*}
Simulating light propagation is central to modern imaging system design \cite{sitzmann_end--end_2018, yang_automatic_2022, wang__2022}. By
modeling how light reflects, refracts, and diffracts through optical elements,
engineers can optimize designs before making physical prototypes.
Simultaneously, computational methods are being augmented with gradient
information, so that first order optimizers can be used in the design process,
greatly reducing the necessary simulation time for systems with large design parameters and enabling end-to-end co-design with machine learning algorithms in a field known as deep optics \cite{zhu_metalens_2023, sun_end--end_2021, li_end--end_2021, yang_automatic_2022, yang_curriculum_2024, sitzmann_end--end_2018, wang__2022}.
These propagation models often rely on simplifying, plane wave assumptions that limit
physical accuracy
in the name of speed. Common approaches are geometric optics (GO)
\cite{pathak_electromagnetic_2022}, physical optics (PO) \cite{pathak_electromagnetic_2022}, and Fourier optics (FO)
\cite{goodman_introduction_2017}. However, as designers push the limits of optical
systems, existing models are increasingly insufficient to model wave effects of desired lens systems.

Our approach enables optimization-based co-design of compound refractive and diffractive optical element (DOE) systems considering both polarized
and unpolarized light (Figure~\ref{fig:overview}), offering distinct advantages over purely refractive systems for applications
requiring miniaturization and broadband performance \cite{geday_adaptive_2020, luo_achromatic_2023}. Unlike refractive elements
that exhibit positive chromatic dispersion, DOEs can be designed for negative dispersion characteristics enabling superior
chromatic correction while reducing overall weight and thickness. Furthermore, systems that capture polarized states of light
have demonstrated superior performance for normal estimation \cite{lei_shape_2022, wang_shape_2025}, reflection reduction \cite{bian_reflection_2024}, and material classification \cite{chen_polarization_1996}; however, due to past solver limitations,
polarized lens design have been underexplored in deep optics.

Previously, simulation of polarized and multiple DOE systems with wave optics were reserved for slow, high-fidelity methods
such at the finite difference time domain (FDTD) \cite{yee_finite-difference_1997}. Though accurate, FDTD simulations
can be prohibitively expensive, on the order of hours to days, especially when considering wide fields of view (FoV). GO can efficiently propagate light for wide FoV but ignores all wave effects \cite{goodman_introduction_2017, balanis_balanis_2024}. PO and FO are able to capture the wave behavior of light, and FO can even be extended to polarized light, but PO cannot model
compound systems and FO is limited to small FoV.

\noindent {\bf Main Results:} In this paper, we present a new lens propagation algorithm using the shooting and bouncing ray (SBR) method \cite{ling_shooting_1989,brem_shooting_2015} to address these issues. SBR combines both GO and PO using the surface equivalence principal \cite{balanis_balanis_2024} to accurately capture many wave effects but without incurring the prohibitive cost of FDTD. The approach grew out of the radio frequency (RF) engineering community as
a high frequency approximation for electromagnetic scattering, but to our knowledge, it has not been used for optimization-based optical engineering before. Our key contributions are: 
\begin{itemize}
  \item A fully differentiable light propagation model based on SBR that
    seamlessly models refractive and diffractive effects across
    all lenses and DOEs in the system. Unlike previous methods we are
    able to characterize polarized light and place no restrictions
    on the location of the DOE.
  \item Validation of the SBR wave-optics method for optical system design by
    comparing against paraxial FO \cite{deb_chromatix_2023, deb_chromatix_2025}, existing ray-wave models \cite{yang_end--end_2024}, and the
    analytical Fraunhofer diffraction patterns \cite{goodman_introduction_2017, richards_electromagnetic_1959} for a
    circular plano-convex lens (polarized and unpolarized) and a
    sinusoidal phase grating. DiffRayve achieves 10 to 30 dB higher peak signal-to-noise ratios than the other baselines compared to analytical solutions.
  \item A
    wide field-of-view compound lens
    systems not representable in FO or past ray-wave models, where
    we are able to halve the chromatic aberration as measured by a weighted mean and variance of the point spread function (PSF). 
\end{itemize}

\section{Related Work}
Numerical simulation is a cornerstone of engineering design optimization.
Zero order methods like Bayesian optimization \cite{jones_efficient_1998} do not require
simulation changes, but performance degrades rapidly as design parameters increase.
Optimizing the phase mask of a DOE, which may contain thousands of parameters, is
often infeasible with these approaches. In contrast, first order methods leverage
simulation gradients to feed algorithms like stochastic gradient descent
(SGD)~\cite{kiefer_stochastic_1952}, enabling fast convergence for high dimensional problems.

Recent lens design research utilizes automatic differentiation frameworks such as
PyTorch~\cite{paszke_pytorch_2019}, JAX~\cite{bradbury_jax_2018}, Zygote~\cite{innes_differentiable_2019},
and Dr. Jit~\cite{jakob_drjit_2022} to facilitate gradient updates.
Sitzmann et al.~\cite{sitzmann_end--end_2018} pioneered end-to-end design using differentiable
Geometric Optics (GO). However, pure GO neglects wave effects crucial for effectively
optimizing optical designs~\cite{ho_differentiable_2025}. Alternatively, differentiable
Fourier Optics (FO) methods~\cite{deb_chromatix_2023, deb_chromatix_2025} model the wave
nature of light but struggle with the paraxial approximations required for large field-of-view
simulations. Consequently, research has shifted toward hybrid methods that model both the
wave and ray nature of light.

Zhu et al. \cite{zhu_metalens_2023} employed a neural surrogate to model metalens wave
propagation followed by GO ray tracing. Although differentiable, this wave-ray method is
bottlenecked by ray conversion costs and limited by surrogate accuracy. Ho et al. \cite{ho_differentiable_2025}
took the inverse approach, propagating fields as GO rays until the final element before
converting to a wave field. This method ignores DOEs and is limited to refractive optics.
Yang et al. \cite{yang_end--end_2024} extended this by adding a learnable phase mask at the
ray-to-wave transition to approximate diffractive effects.

While previous methods combine ray and wave models, they isolate them to distinct regions
of the system. This rigidity restricts where DOEs can be placed. FO handles diffractive
elements globally but fails at wide fields of view \cite{goodman_introduction_2017}. GO
handles wide fields of view but ignores diffraction. Hybrids like DeepLens apply a phase
offset only at the final element, forcing DOEs to the end of the system. Furthermore,
current hybrid methods \cite{zhu_metalens_2023, yang_end--end_2024, ho_differentiable_2025}
cannot model polarized light. We summarize these comparisons in Table~\ref{tab:perf-comp}.
By building our simulation on SBR, we consider ray and wave models simultaneously to
address these limitations.
\begin{table}[t]
  \centering
  \setlength{\tabcolsep}{3.0pt} 
  \caption{\textbf{Feature Comparison.} Comparison of lens simulation
  models (Fourier Optics, Geometric Optics, DeepLens, and our method) for combination diffractive and refractive systems.}
  \label{tab:perf-comp}
  \begin{tabular}{l|ccccc}
    \toprule
    Features & FO \cite{deb_chromatix_2023} & GO \cite{pathak_electromagnetic_2022} &
    DeepLens \cite{yang_curriculum_2024} & Ours \\
    \midrule
    Wide FoV & \xmark & \cmark & \cmark &  \cmark \\
    Diffraction & \cmark & \xmark & \cmark & \cmark \\
    Polarization & \cmark & \xmark & \xmark & \cmark \\
    Multiple DoEs & \cmark & \xmark & \xmark & \cmark \\
    \bottomrule
  \end{tabular}
\end{table}
\section{Shooting and Bouncing Ray Method}
\label{sec:background}
We provide a short description of the shooting and bouncing ray method. For in-depth
details about the method and concepts in electromagnetics we turn readers
to~\cite{brem_shooting_2015, ling_shooting_1989, balanis_balanis_2024, audia_accelerated_2025}.

{\bf Notation:} Throughout, values in bold represent vectors even if they are capitalized. We use
this convention so that the current and field vector match their typical
description in electrical engineering texts. $\hat{\mathbf{n}}$ is
the unit surface normal. We also use multiple constants common to
electromagnetics. These are
the impedance of free space $\eta_0$, permittivity of free space
$\epsilon_0$, and permeability of free space $\mu_0$. The wave
number is $k_0 = \frac{2\pi}{\lambda}$, where $\lambda$ is the
wavelength for a given
frequency in the propagating medium.
A factor of $e^{j\omega t}$ is assumed throughout, where $t$ is time and $\omega$ is angular frequency.

\begin{figure}[t]
  \centering
  \includegraphics[width=1.0\linewidth]{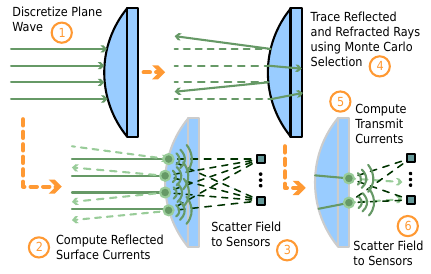}
  \caption{\textbf{Overview of Our Ray-Wave Algorithm.} 1. Discretize and trace
    a plane wave of GO rays. 2. Compute the equivalent surface currents
    using local GO field. 3. Scatter and sum the field from each surface
    current at each sensor pixel. 4. Trace another set of rays randomly
    selecting reflected and transmit directions based on material properties.
    5. Compute the next set of surface currents. 6. Scatter these currents
    to the pixel sensors again. These two stages repeat until rays have
  left the system.}
  \label{fig:sbr-steps}
\end{figure}

The shooting and bouncing ray method~\cite{ling_shooting_1989, brem_shooting_2015, audia_accelerated_2025} combines multiple techniques to calculate
an efficient and accurate approximation of the electric field. The method iterates three key steps:
propagation of GO waves, computation of PO currents at the surface, and scattering of the currents to the sensor using the electric field integral equation (EFIE) \cite{balanis_balanis_2024} (see Figure~\ref{fig:sbr-steps}). The field enters
the aperture as a plane wave discretization as GO rays, each described by 
\begin{equation}
    \mathbf{E} = \mathbf{E}_0e^{-jk_0\mathbf{\hat{k}}\cdot\mathbf{r}},
\end{equation}
where $\mathbf{E}_0$ is the initial field intensity and polarization, $\mathbf{\hat{k}}$ is
the propagation direction, and $\mathbf{r}$ is the spatial coordinate. The orthogonal magnetic field, $\mathbf{H}$, 
is computed by $\mathbf{H} = \frac{1}{\eta}\mathbf{\hat{k}}\times\mathbf{E}$. GO rays are traced
throughout the geometry, tracking both phase and polarization information influenced
by changing indices of refraction and Fresnel coefficients \cite{balanis_balanis_2024}.

At each surface intersection, equivalent electric and magnetic currents densities are computed using $\mathbf{J} = \mathbf{\hat{n}}\times\mathbf{H}$
and $\mathbf{M}=\mathbf{E}\times\mathbf{\hat{n}}$, respectively.
Currents are always computed on the outermost surface, where the currents are "visible" to the sensor. 
When a field enters a lens, the electric and magnetic fields are approximated by the incident GO wave and
its reflection. Alternatively, when the field leaves the lens as it refracts, only the transmitted ray
field is considered when computing the current. In contrast to other hybrid wave-optics methods,
equivalent currents are computed along all optical element boundaries, not just after
the final element, allowing for SBR to characterize multiple DOEs.

Given $\mathbf{J}$ and $\mathbf{M}$, the scattered electric field
generated at a point
$\mathbf{r}$ is given by
the EFIE,
\begin{multline}
  \mathbf{E}^s(\mathbf{r}) = jk_0\eta_0\iint_\Gamma (\mathbf{I} +
  \frac{\nabla\nabla}{k_0^2})G(\mathbf{r}, \mathbf{r}')\cdot
  \mathbf{J}(\mathbf{r}') \\
  + \frac{1}{\eta_0} \mathbf{M}(\mathbf{r}') \times \nabla
  G(\mathbf{r}, \mathbf{r}')d\mathbf{r}',
  \label{eq:efie}
\end{multline}
where $G(\mathbf{r}, \mathbf{r}')=\frac{e^{-jk_0|\mathbf{r}-\mathbf{r}'|}}{4\pi
|\mathbf{r}-\mathbf{r}'|}$ is the scalar Green's function
for the 3D Helmholtz equation and $\Gamma$ is the boundary between two homogeneous, isotropic regions. The location of the surface current density is $\mathbf{r}^\prime$. 
\section{DiffRayve: Differentiable Ray-Wave}
\label{sec:method}
We now adapt the shooting and bouncing ray algorithm
presented above to formulate our method, DiffRayve, by introducing near field scattering and automatic differentiation of 
the Monte Carlo accelerated SBR algorithm. Additional details about the implementation can be found in Appendix~A.
\subsection{SBR Simplification of EFIE}
The EFIE as it stands in Equation~\ref{eq:efie} contains multiple singularities and derivatives
that are difficult to integrate. Most SBR implementations consider scattered fields in the far-field,
where the resulting field is also a plane wave. In lens systems, this approximation is not sufficient
and instead we must consider the radiative near-field form of the equation as
\begin{multline}
  \mathbf{E}^s(\mathbf{r}) \approx
  jk_0\eta_0\iint_\Gamma
  (\hat{\mathbf{r}}\times\hat{\mathbf{r}}\times
    \mathbf{J}(\mathbf{r}')
    + \\
  \frac{1}{\eta_0}\hat{\mathbf{r}}\times\mathbf{M}(\mathbf{r}'))\frac{e^{-jk_0|\mathbf{r}-\mathbf{r}'|}}{4\pi|\mathbf{r}-\mathbf{r}'|}d\mathbf{r}',
  \label{eq:efie-near-field}
\end{multline}
where $\hat{\mathbf{r}}$ is the unit vector pointing from
${\mathbf{r}}'$ to ${\mathbf{r}}$. We give further details about this simplification in Appendix~A.1.
\subsection{Monte Carlo Near-Field}
We adapt the work of Audia et al. \cite{audia_accelerated_2025} to Monte Carlo integrate the
near-field integral presented above. As demonstrated in that work, the computational
and memory complexity of the SBR algorithm can be made linear rather than exponential
in the number of optical elements. We first split Equation~\ref{eq:efie-near-field}
into paths of the same length. For example, paths that intersect the front of the first
optical element (step 1 in Figure~\ref{fig:sbr-steps}), have a length of two, where
the first point was on the descretized plane wave. As subsequent rays are traced and intersected
with the lens elements, additional points are added to each path. The scattered field
for a given path length, $N$, is expressed by
\begin{multline}
  \mathbf{E}^s_N(\mathbf{r}) = jk_0\eta_0
  \overbrace{\sum^{2} \cdots \sum^{2}}^{N-2}
  \int_{\Gamma_1} \\
  \frac{(\mathbf{\hat{r}} \times \mathbf{\hat{r}} \times \mathbf{J}(\mathbf{p}) +
      \frac{1}{\eta_0}\mathbf{\hat{r}}\times
  \mathbf{M}(\mathbf{p}))}
  {|\mathbf{\hat{k}}^r_N \cdot \mathbf{\hat{n}}(\mathbf{r}^\prime)|}
  \frac{e^{-jk_0|\mathbf{r}-\mathbf{r}'|}}{4\pi|\mathbf{r}-\mathbf{r}'|}
  dA(\mathbf{p}_1),
  \label{eq:scattering-path-space}
\end{multline}
where $\Gamma_1$ is the cross sectional area of the incident plane wave
given by the bounding box of the lens system for a given incident direction
with the initial starting point $\mathbf{p}_1$.
$p$ is the full set of points that describes a path built by multiple ray
traces and $\mathbf{\hat{k}}^r$ is the unit direction between the second
to last point and the last point on the path, $\mathbf{r}^\prime$. The discrete
sums are produced by choosing either the reflected or refracted path. To fully characterize
the field, the paths of all lengths are considered such that the total scattered
field is $\mathbf{E}(\mathbf{r}) = \sum_{l=2}^\infty \mathbf{E}^s_l(\mathbf{r})$.

Equation~\ref{eq:scattering-path-space} is then approximated using Monte Carlo integration
with $N_s$ samples and paths, $\mathbf{p}_i$, indexed by $i$ to produce
\begin{multline}
  \mathbf{E}^s_N(\mathbf{r}) \approx jk_0\eta_0\frac{\int_{\Gamma_1}dA(\mathbf{p}_1)}{N_s} \\ 
  \sum_{i=1}^{N_s} 
  \frac{(\mathbf{\hat{r}}_i \times \mathbf{\hat{r}}_i \times \mathbf{J}(\mathbf{p}_i) +
      \frac{1}{\eta_0}\mathbf{\hat{r}}_i\times
  \mathbf{M}(\mathbf{p}_i))}
  {p(\mathbf{p}_i)|\mathbf{\hat{k}}^r_{N,i} \cdot \mathbf{\hat{n}}(\mathbf{r}^\prime_i)|}
  \frac{e^{-jk_0|\mathbf{r}-\mathbf{r}'_i|}}{4\pi|\mathbf{r}-\mathbf{r}'_i|},
  \label{eq:scattering-path-space-mci}
\end{multline}
where $p(\mathbf{p}_i) = \prod_{l=1}^N(\mathbf{p}_l)$ is the probability of a given
path defined as the product of probabilities of each ray sample. Rays are initially 
sampled from the incident plane wave using stratified sampling \cite{pharr_physically_2016}
which subdivides the plane into uniform strata and uniformly samples from each strata. 
This method allows for proper sampling of field oscillations. We found that a strata
sized to the simulation wavelength of light was a good balance of performance to accuracy.
This parameter is explored in Appendix~A.3. At each lens interaction,
the probability of picking the reflected or refracted direction is weighted by the 
average Fresnel reflection coefficient of each polarization to favor paths with
the most energy. Lastly, as mentioned, the total field is a sum of paths of infinite length.
For implementation, we truncate this path to be two times the number of optical elements
to ensure that all surfaces are intersected at least once (further experiments in Appendix~A.4.
\subsection{Differentiable Shooting and Bouncing Ray}
Differentiably propagating GO rays is well established and
handled by libraries like Dr. Jit
\cite{jakob_drjit_2022} and Mitsuba \cite{jakob_mitsuba_2022}, on which
we implemented our algorithm. Please note that while Mitsuba already
implements many path tracing algorithms, it does not implement
SBR, and we leveraged these libraries only for their GPU acceleration,
automatic differentiation, and basic ray tracing optimizations. For our work,
we compute gradients by applying automatic differentiation directly to
Equation~\ref{eq:scattering-path-space-mci}. There are known limitations
to this approach \cite{vicini_path_2021}; however, due to the low dimensionality
of light paths and limited discontinuities in optimizing DOEs, we find that 
this approach achieves strong results in lens design applications as demonstrated below.
\subsection{Polarized and Unpolarized Light}
To serve as a general optical system solver, we must compute either polarized or unpolarized light. To do so, we assume that two
orthogonal incident polarizations, $V$ and $H$, are incoherent. In the end, we have two scattered fields $\mathbf{E}^V$ and
$\mathbf{E}^H$. The final
field intensity at each sensor pixel location is then calculated by
\begin{equation}
  I(\mathbf{r}) = \frac{1}{2} (|\mathbf{E}^V(\mathbf{r})|^2 +
  |\mathbf{E}^H(\mathbf{r})|^2).
  \label{eq:unpolarized-incoherent}
\end{equation}
\subsection{Diffractive-Refractive Element Models}
In or implementation, both refractive and diffractive
elements are modeled using 3D triangular meshes. A texture
is applied to the mesh to map the relative permeability $\mu_r$, and
relative permittivity
$\epsilon_r$, for the inner
and outer regions with respect to the right-handed triangle normal. Diffractive
elements use an additional intermediate representation for
accumulating gradients.
Two element types are considered. First, a pixel height map uses a 2D texture
to calculated positive and negative offsets of the diffractive surface. Triangle
vertices are then adjusted by the texture offset before ray tracing
in each iteration of the gradient descent algorithm.
The second model is a binary phase mask represented by $\sum_{i=1}^5
\alpha_i \rho^{2i}$,
where $\alpha_i$ are learnable coefficients and $\rho$ is the 2D distance from
the central axis. This representation was chosen to work well with
existing features
of Mitsuba. 
\subsection{Sensor Model}
In the present work we use a simple sensor
model; however, this not a limitation of the current work. New sensors could be included by post processing the calculated electric field using a differentiable framework as was done in \cite{sun_end--end_2021}. We assume that pixels are
located in a rectangular grid defined by a width $w$, and height
$h$, in meters.
The pixel counts along each dimension are given by the integers $N_w$ and $N_h$.
The sensor is oriented by a point $\mathbf{p}$ and two position vectors
$\hat{\mathbf{f}}$ and $\hat{\mathbf{u}}$ for forward and up.
$\hat{\mathbf{u}}$ aligns along
the height dimension of the sensor. The vector $\hat{\mathbf{r}}$ aligned with
the width is found by normalizing the cross product between $\hat{\mathbf{f}}$
and $\hat{\mathbf{u}}$. The location of each pixel indexed by $i$ and
$j$, $\mathbf{P}(i, j)$, is then
calculated using
\begin{equation}
  \mathbf{P}(N_w, N_h) = \mathbf{p} +
  (-\frac{w}{2}+\frac{i}{N_w}w)\hat{\mathbf{f}}
  + (-\frac{h}{2}+\frac{j}{N_h}h)\hat{\mathbf{u}}.
  \label{eq:pixel-locations}
\end{equation}

\section{Results}
\label{sec:results}
In this section, we validate our use of SBR for lens design by comparing against
existing differentiable lens design simulations and analytical
solutions. We then
demonstrate our method's ability to optimize a wide field-of-view system
with multiple DOEs and train a polarization aware lens and classification algorithm end-to-end. Existing state-of-the-art (SOTA) methods
are unable to optimize this system due to limitations in their forward
simulations either making paraxial approximations or limiting the placement of DOEs.
Our experiments were performed on a NVIDIA RTX A6000
with 48 GB of video memory.
\subsection{Validation}
We compare our SBR results with Fourier Optics using Chromatix
\cite{deb_chromatix_2023, deb_chromatix_2025}, a SOTA
ray-wave method \cite{yang_curriculum_2024}, the GO solution, and the
analytical solution, which serves as ground truth. We plot the
normalized point spread function \cite{goodman_introduction_2017} of $550 nm$ light.
Throughout, we find that the SBR algorithm matches strongly against
the analytical Fraunhofer diffraction solution demonstrating that our approach is well suited for optical lens design. Additional details can be found in Appendix~D, including analytical formulas.
\subsubsection{Scalar Wave Optics Scattering}
\begin{figure*}[t]
  \centering
  \includegraphics[width=1.0\textwidth]{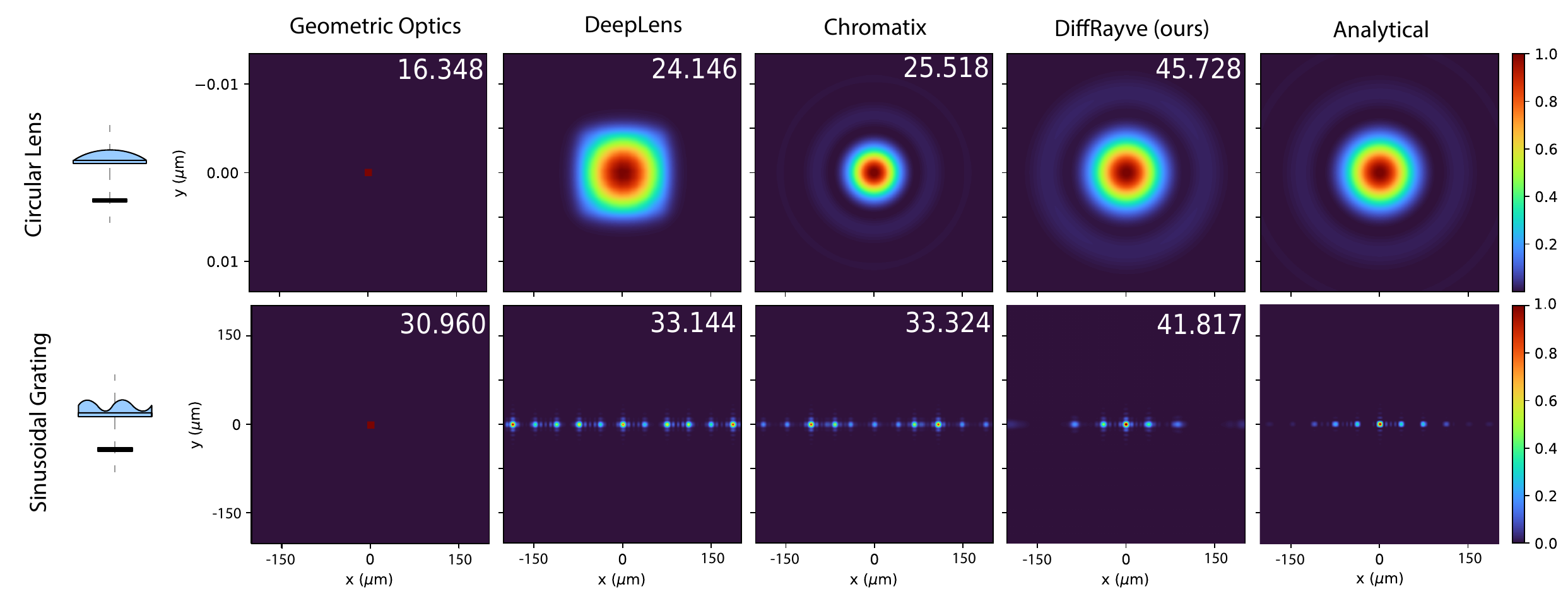}
  \caption{\textbf{Scalar Wave Comparison.} We validate the SBR method for use in optical lens design by comparing against Chromatix's \cite{deb_chromatix_2023, deb_chromatix_2025} paraxial FO implementation and
    state-of-the-art ray-wave method, DeepLens \cite{yang_end--end_2024}, as well as the analytical solution.
    GO \cite{pathak_electromagnetic_2022} is not able represent any diffraction as expected. We include PSNR values relative to the analytical solution are included in each plot. Our method either matches or beats existing approaches and succeeds at modeling the wave nature of light.}
  \label{fig:scalar-wave-comp}
\end{figure*}
Scalar wave optics scattering is explored for both a circular lens and
a square sinusoidal phase grating.  Figure \ref{fig:scalar-wave-comp} compares each method against the
analytical solution qualitatively and quantitatively using peak signal-to-noise ratio (PSNR). We demonstrate that all methods are able to
accurately capture the wave effects of light scattering except for
geometric optics, which predicts only a coalesced point. For the circular lens, our method is 20 dB higher than Chromatix and DeepLens.  
For the sinusoidal grating, our method has a PSNR almost 10 dB higher than DeepLens and Chromatix. Other methods do not fully capture the physics
of the lens system as the paraxial approximation is broken \cite{goodman_introduction_2017}. First, for large 
sensors, the field drops off with a factor of $\cos^4(\theta)$, where $\theta$ is the
horizontal angle between the lens axis and the location on the sensor. Similarly,
the field smears at the corners, elongating the bright spots further from the axis. This effect is due to the spherical expansion of surface currents being projected
on the flat sensor plane. So, while other methods can capture the Fraunhofer field
accurately, out method is closer to what a real world system would measure.
All methods completed their circular lens simulation in under 5 seconds using around $500MB$ of memory. For the sinusoidal grating, DiffRayve completed the simulation in 2.5 seconds and used $220MB$ of memory. Chromatix and DeepLens completed the same simulation in around 1 minute.
\subsubsection{Polarized Wave Optics Scattering}
\begin{figure}[t]
  \begin{center}
    \includegraphics[width=1.0\linewidth]{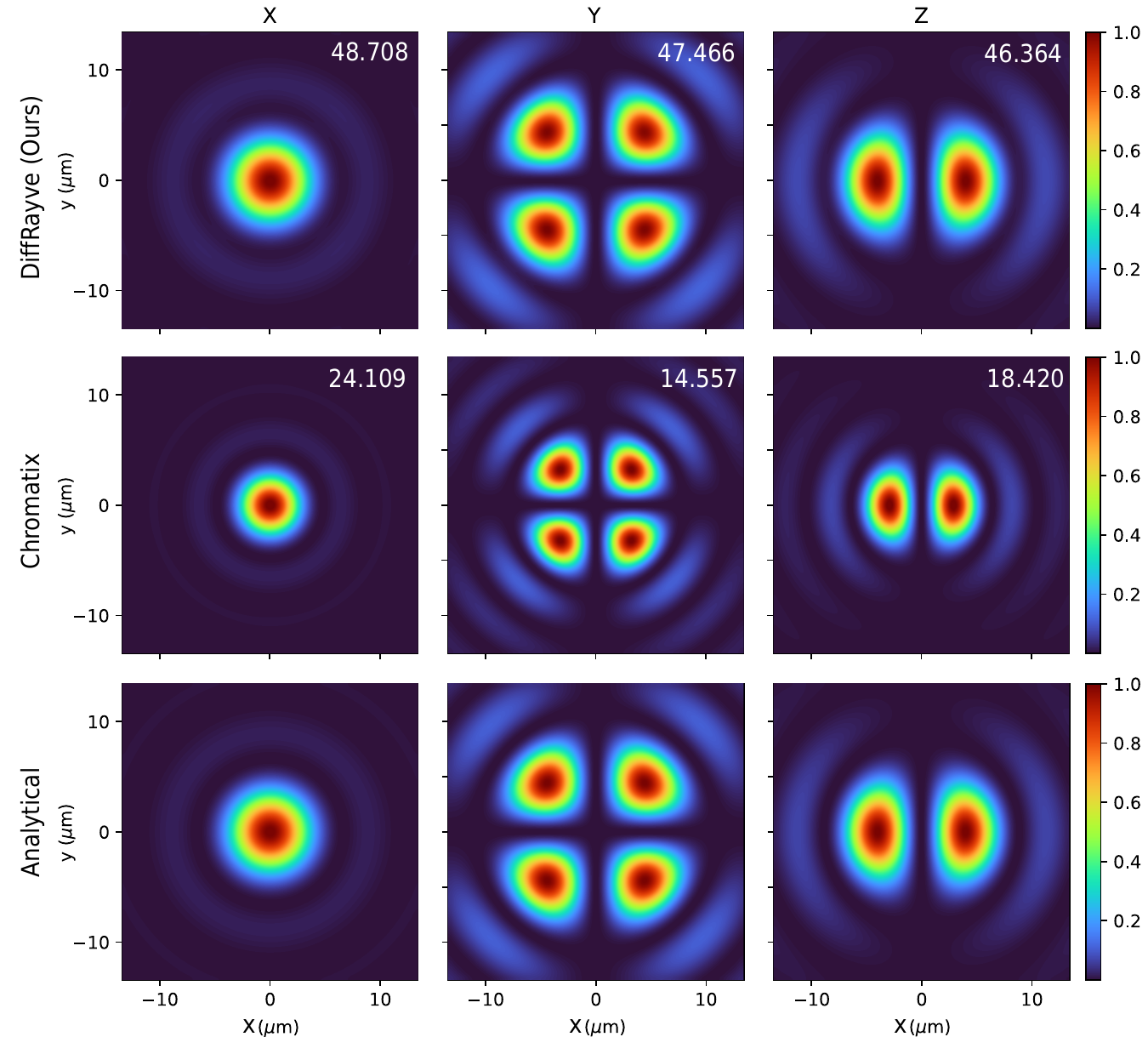}
  \end{center}
  \caption{\textbf{Vector Wave Comparison.} We compare Chromatix's \cite{deb_chromatix_2023, deb_chromatix_2025} paraxial FO, DiffRavye (ours), and
    the analytic solution for a polarized field passing through a circular
    plano-convex lens. PSNR as compared to the analytical solution is displayed in each square. Though Chromatix has higher values, our method is able to capture the structure of the polarized field, further validating the method for polarized lens systems. }\label{fig:vector-wave-comp}
\end{figure}
Our method is also able to characterize the propagation
of polarized light. This is a key advantage of our method
as existing SOTA ray-wave methods \cite{yang_end--end_2024, ho_differentiable_2025, zhu_differentiable_2020} are only able to capture
unpolarized fields. We compare against
the analytical solution and Chromatix \cite{deb_chromatix_2023, deb_chromatix_2025}. Figure \ref{fig:vector-wave-comp} shows
that both methods qualitatively match the analytical solution \cite{richards_electromagnetic_1959} when computing
the polarized field on the other side of a circular lens.
Quantitatively, our method achieves as PSNR 30 dB higher than Chromatix
as compared to the analytical solution. 
\subsection{Multiple Diffractive Element Optimization}
\label{sec:mult-doe-result}
\begin{figure}[t]
    \centering
    \begin{subfigure}[t]{0.49\textwidth}
        \includegraphics[width=\linewidth]{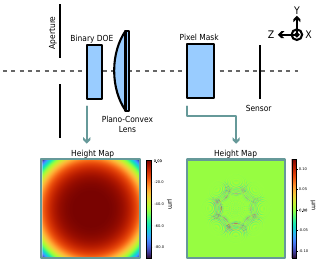}
        \caption{\textbf{Multiple DOE Lens}}
        \label{fig:multi-doe-optimized}
    \end{subfigure}
    \begin{subfigure}[t]{0.49\textwidth}
        \centering
        \includegraphics[width=\linewidth]{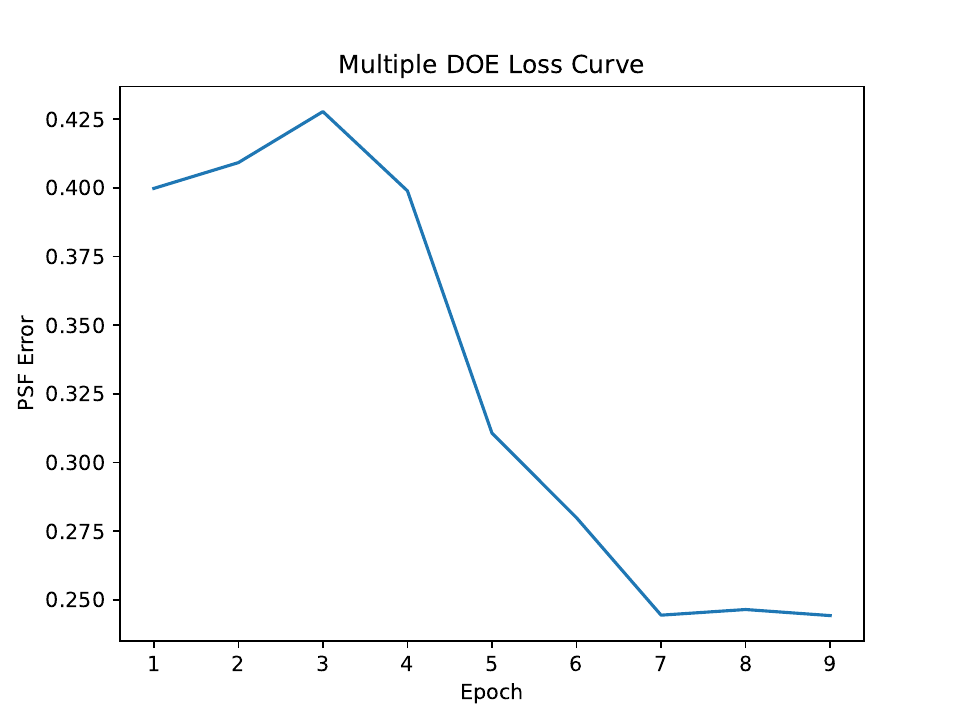}
      \caption{\textbf{Multiple DOE Loss Curve}}
      \label{fig:multi-doe-loss}
    \end{subfigure}
    \caption{\textbf{Multiple DOE Lens Optimization.} This figure depicts the lens system used in the optimization experiment (Section~\ref{sec:mult-doe-result}). For detailed dimensions of the system please see that section. 
    The system contains three elements: a binary DOE, a plano-convex lens, and a pixel mask DOE. The height maps for the optimized diffractive elements are shown. The pixel mask shows a repeating circular pattern that focuses on the light on the sensor.}
\end{figure}

We demonstrate a key benefit of our work by designing a wide
field-of-view system with multiple diffractive optical elements.
From aperture to sensor, our system begins with a $1mm$ diameter circular aperture,
a $2 mm$ binary phase element, then a $3 mm$ diameter
plano-convex refractive lens with a radius of curvature of $2.5 mm$, and lastly, a $2 mm$ pixel diffractive phase mask. The binary DOE has a 
max thickness of $1mm$ with its center $1mm$ behind the aperture. 
The refractive lens has a thickness of $0.5 mm$ and it front is $1.5$
millimeters past the aperture. The last pixel diffractive mask has a max
thickness of $2 mm$ and is $2 mm$ in front of the imaging sensor, which
is placed at the focal length of the refractive lens. All elements are given
a refractive index of 1.5. The layout is shown in Figure~\ref{fig:multi-doe-optimized}. 
The binary DOE has 5 learnable parameters while the pixel mask
has $2^{14}$. The optimization used $28.5 GB$ of GPU memory and took 10 minutes. For more details, please see Appendix~B.
\begin{figure}[t!]
  \centering
  \begin{subfigure}[t]{1.0\columnwidth}
    \centering
    \includegraphics[width=\linewidth]{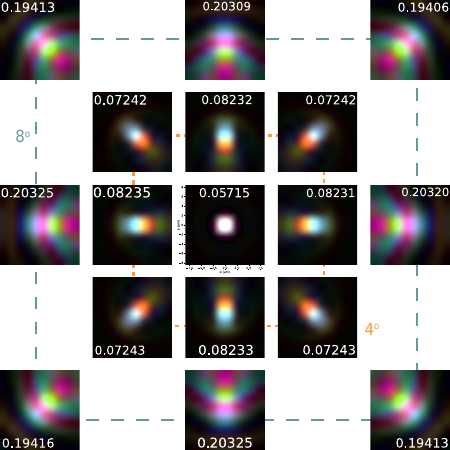}
    \caption{Starting PSFs}
    \label{fig:mult-doe-pre}
  \end{subfigure}
  \hfill
  \begin{subfigure}[t]{1.0\columnwidth}
    \centering
    \includegraphics[width=\linewidth]{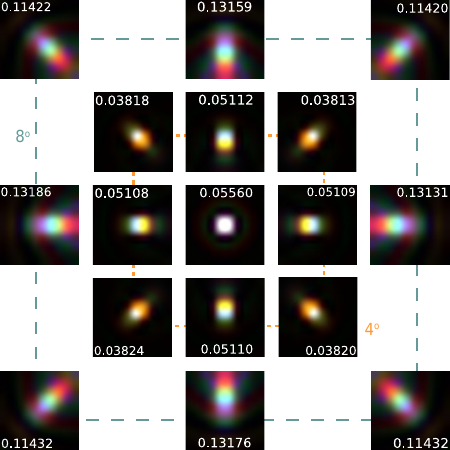}
    \caption{Optimized PSFs}
    \label{fig:mult-doe-post}
  \end{subfigure}
  \caption{\textbf{Optimized Multiple DOE PSFs.} The lens system is
    able to focus the
    PSF for multiple wavelengths and wide fields of view. Each square is a $10 \times 10$ grid with a pixel pitch of $1.6 \mu m$. The overall dimensions of each square is then $16 \mu m$. The PSFs are placed in their relative location on the sensor plane as focused by the lens system. The error (Equation~\ref{eq:mult-doe-loss}) associated with each square is the loss function average across the three color channels. We see that the optimized DOEs produces PSFs that are condensed and closer to a white dot reducing chromatic aberration.
    }
  \label{fig:multi-doe-psfs}
\end{figure}

We used a loss function, $L(\mathbf{P}_k)$ that penalizes the size and location of the $k^{th}$ PSF, $\mathbf{P}_k$, of the training batch by summing a weighted mean and variance, given by
\begin{equation}
    L = \frac{1}{16}\sum_{k=1}^{16\cdot3}\Big[\sigma^2_k+\frac{1}{100}\sum_{i=1}^{10}\sum_{j=1}^{10}\rho_{i,j}\hat{p}^{(k)}_{i,j}\Big].
    \label{eq:mult-doe-loss}
\end{equation}
Before optimization, the PSFs display a large
amount of chromatic aberration. The $8^\circ$
incident angles need the most correction as their field 
is spread out over the whole grid (Figure~\ref{fig:mult-doe-pre}). After optimization (Figure~\ref{fig:mult-doe-post}),
the PSFs are condensed to the center of the square converging
on a white dot where all three color channels align. Quantitively, we see a nearly a halving of error on the $8^\circ$ PSFs and a $66 \%$ decrease in error at $4^\circ$.
The paraxial PSF saw little improvement, but overall the systems error dropped from 0.4 to 0.25 (Figure~\ref{fig:multi-doe-loss}).

Though initialized to a flat
height map,
the pixel phase mask DOE (Figure~\ref{fig:multi-doe-optimized}) lens element appears similar to multiple Kinoform lenses
\cite{jordan_kinoform_1970}, centered at angles of interest.
The loss curve in Figure \ref{fig:multi-doe-loss} shows that gradients from
the simulation quickly achieve the optimization objective. Appendix~C
demonstrates that these improved PSFs hold for angles not in the training dataset.
We also demonstrate the ability to optimize a vector field in Appendix~E.
\section{Conclusion, Limitations, and Future Work}
Our work introduces a novel differentiable method for optical lens design,
inspired by the shooting and bouncing ray (SBR) approach commonly used in RF
design. By integrating the strengths of geometric and physical optics, our
method accurately models refractive, diffractive, and polarized light
behavior. We
demonstrate our method's effectiveness by optimizing a multiple DOE
system
 for wide field-of-view imaging. Our
approach has broad applications as a
general solver for high-fidelity optical systems and paves the way for the
design of new, complex lenses. We also demonstrate our solvers use in end-to-end training by co-designing a lens and classification network leveraging polarized scattering.

Our approach has some limitations. To calculate unpolarized systems, DiffRayve must run two ray traces to compute a single wavelength's PSF. These can be combined into a single ray trace because both polarizations follow the same GO paths. If the material properties are also assumed to be the same for all wavelengths, a single ray trace can be performed for all polarizations and color channels. The wavelength is only different when tracking phase and scattering to the sensor. Until such
optimizations are realized,
simpler refractive systems may be better served by existing solvers.
Similarly, SBR's approximations
on the surface currents mean that it cannot characterize discontinuities on elements that are on the order of a wavelength. Therefore, meta-lens systems \cite{hazineh_d-flat_2022} may still require higher fidelity methods.
Nonetheless, our work opens new possibilities for lens design, enabling
engineers to co-design systems that account for both the wave and ray nature of
light.

{
    \small
    \bibliographystyle{ieeenat_fullname}
    \bibliography{references}
}

\appendix
\section{DiffRayve Details}
\label{sec:diffrayve-extra-details}

We briefly discuss implementation details that were
used in DiffRayve. We demonstrate that the decisions 
used do not influence the results in the main paper. We will pursue open sourcing our simulation upon acceptance. 

\subsection{Near Field EFIE}
\label{sec:near-field-integral-details}

\begin{figure}[h]
  \centering
  \includegraphics[width=0.85\linewidth]{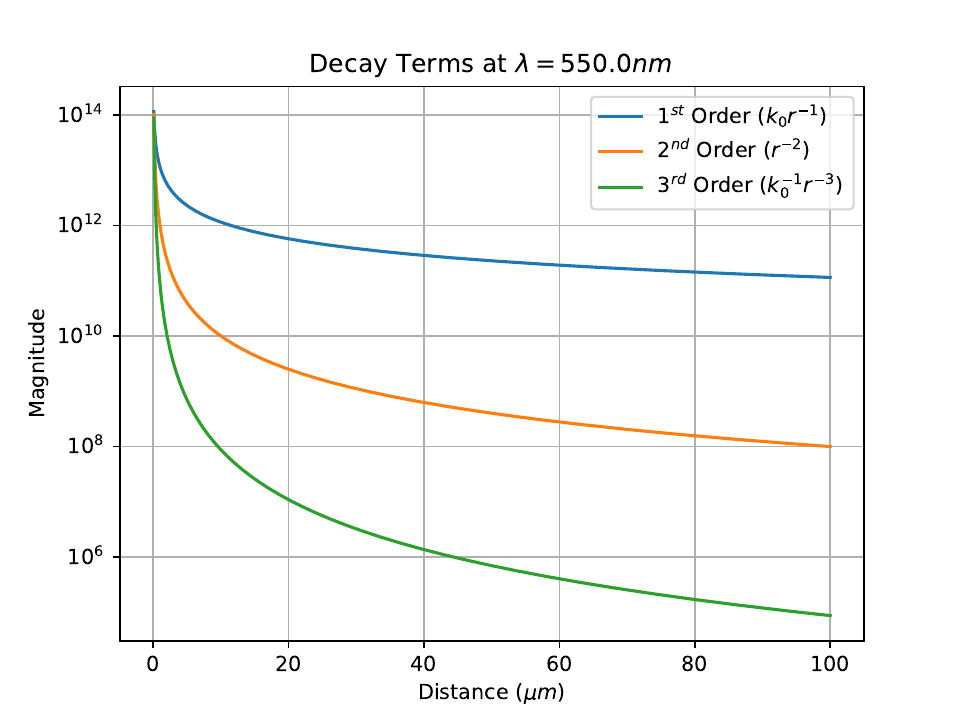}
  \caption{\textbf{Higher Order Terms.} We justify our simplification of Equation~2 by showing that higher order terms decrease faster than the leading term at the wavelengths and scales typical in lens design. At $\lambda = 550nm$ and $100 \mu m$, the $k_0r^{-1}$
    term is 3 and 6 orders of
    magnitude larger than the $r^{-2}$ and $k_0^{-1}r^{-3}$,
    respectively.}
  \label{fig:higher-order-terms}
\end{figure}

We begin our derivation by simplifying the EFIE (Equation~2) specifically for lens design applications.
Typically, SBR simplifies the EFIE to far-field plane waves
by only considering the leading order terms and truncating the Taylor
expansion of exponential terms. If this equation was used for lens design,
then there would be no difference in field along the sensor. We can improve the fidelity by considering additional terms. To determine which higher order terms to include, we consider a surrogate
problem of an infinitesimal dipole current, where Equation 2
can be computed in closed form \cite{balanis_balanis_2024}. Three singular
terms are proportional to $k_0r^{-1}$, $r^{-2}$, and $k_0^{-1}r^{-3}$. These
terms are plotted in Figure \ref{fig:higher-order-terms} for a 550 nanometer
wavelength. At a distance of $100 \mu m$, the inverse square and inverse
cubic terms are multiple orders of magnitude lower than the leading term.
As such, we keep only the inverse term and use the radiative near-field integral
for propagation to the sensor, giving

\subsection{Scaling}

\begin{figure}[h]
    \centering
    \includegraphics[width=0.75\linewidth]{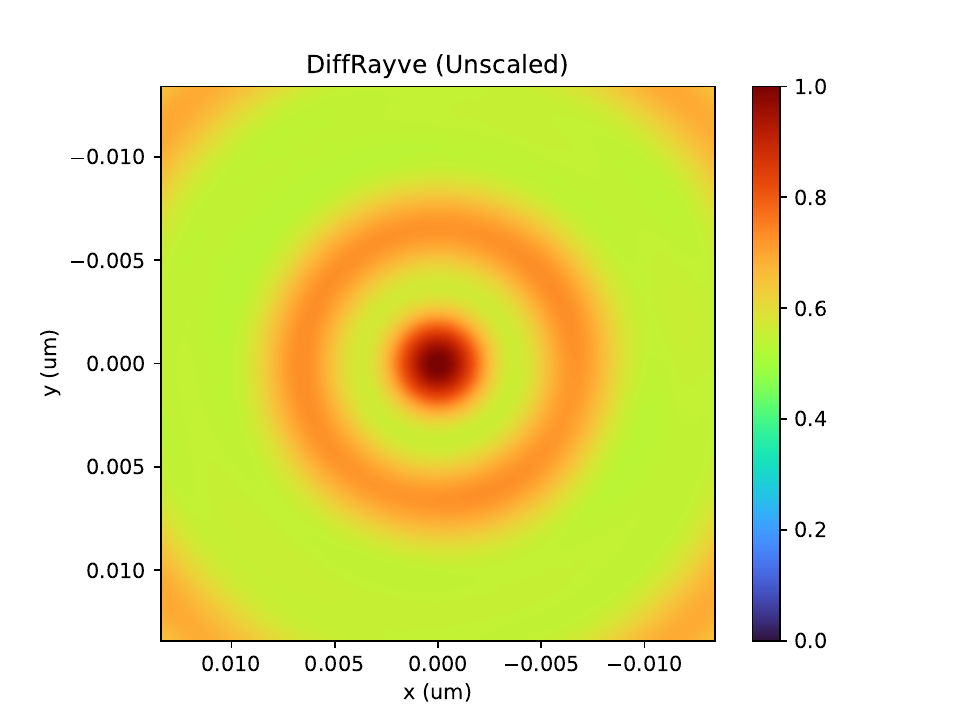}
    \caption{Lens results with no scaling factor applied causing rays to leak through the lens.}
    \label{fig:unscaled-result}
\end{figure}

In the original implementation, we found that there were
numerical issues when running a 3 millimeter lens at 550 nanometer (Figure \ref{fig:unscaled-result}). These parameters resulted in a loss 
of detail in the computed field. One potential explanation are internal operations in Mitsuba \cite{jakob_mitsuba_2022} not being able to handle self intersections properly at these scales. To mitigate the problem, we scaled
the system so that the input dimensions where in millimeters (i.e
$1e{-3} \rightarrow 1$). This is valid as long as the frequency is lowered by the same factor, resulting in a wavelength on the order of $5.5e{-4}$. 
This is a common approach in chamber measurements to allow for smaller
facilities \cite{sinclair_theory_1948}.

Assuming a scale factor $p$, we see that the phase factor is updated by
\begin{equation}
    e^{-jk_0r} = e^{-j\frac{2\pi f}{c_0}r} \rightarrow e^{-j\frac{2\pi f}{p c_0}rp} = e^{-j\frac{2\pi f}{c_0}r}.
\end{equation}
We see that the scaling terms cancel so that the constructive and destructive interference of individual currents are the same. The 
scaling does result in an overall scaling of the scattered
field by a factor of $\frac{1}{p}$; however, when computing point spread functions, we normalize the field removing the influence of this scale term. Even without normalization the equivalent field intensity can
be found by multiplying by $p$. After scaling, we find that the field 
has greater dynamic range as expected (see Figure in main paper).

\subsection{Monte Carlo Sampling Probabilities}

In Monte Carlo Shooting and Bouncing Rays (SBR) \cite{audia_accelerated_2025}, the reflected and refracted ray directions can be sampled using different probability
distributions. We follow the original work and weight the decision
based on the average reflection coefficients. At near normal incidence,
this weights rays towards refraction so that all lens elements are 
sampled as transmission coefficients are larger than reflection coefficients.

\subsection{Bounce Count}
\label{sec:bounce-count}

\begin{figure}[h]
  \centering
  \begin{subfigure}[b]{0.19\columnwidth}
    \centering
    \includegraphics[width=\linewidth]{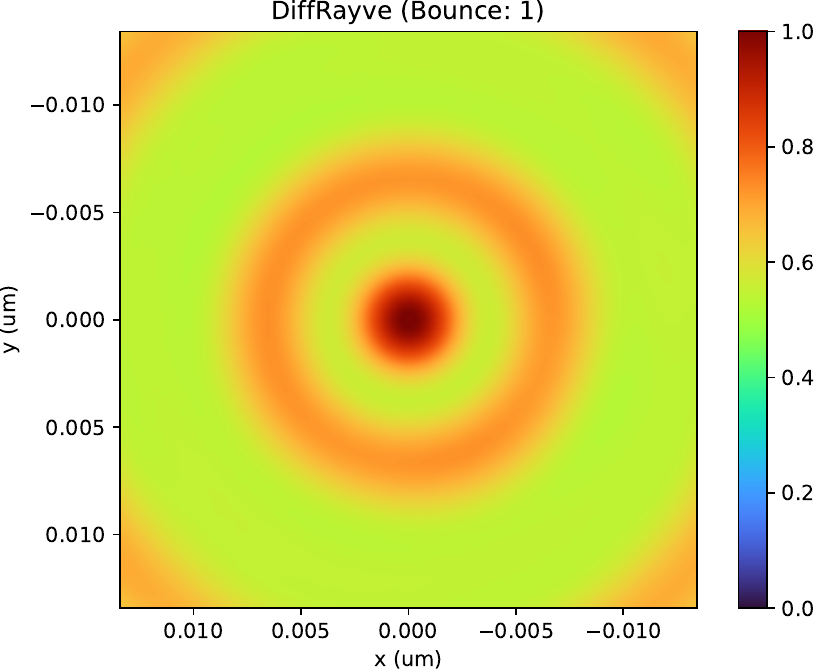}
    \caption{1 Bounce}
  \end{subfigure}
  \hfill
  \begin{subfigure}[b]{0.19\columnwidth}
    \centering
    \includegraphics[width=\linewidth]{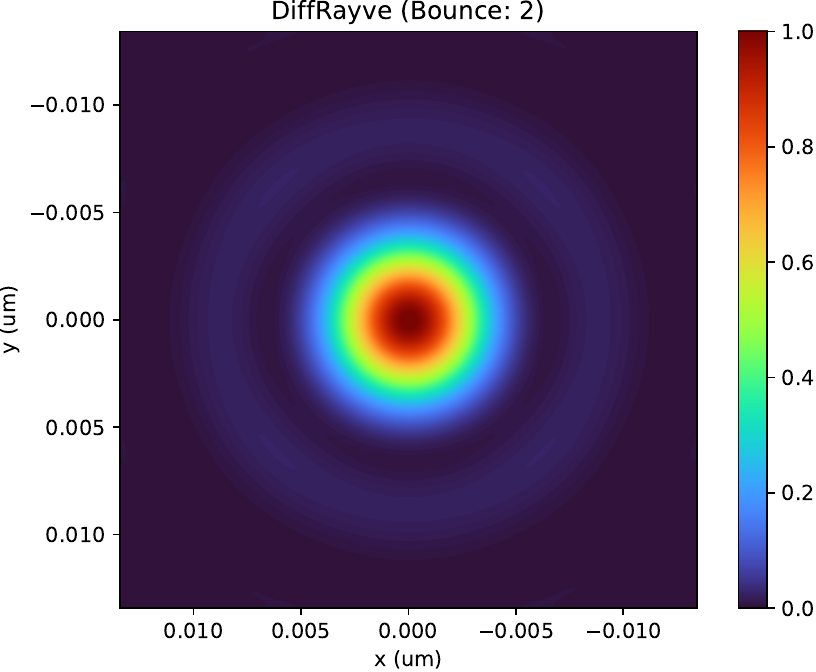}
    \caption{2 Bounces}
  \end{subfigure}
  \centering
  \begin{subfigure}[b]{0.19\columnwidth}
    \centering
    \includegraphics[width=\linewidth]{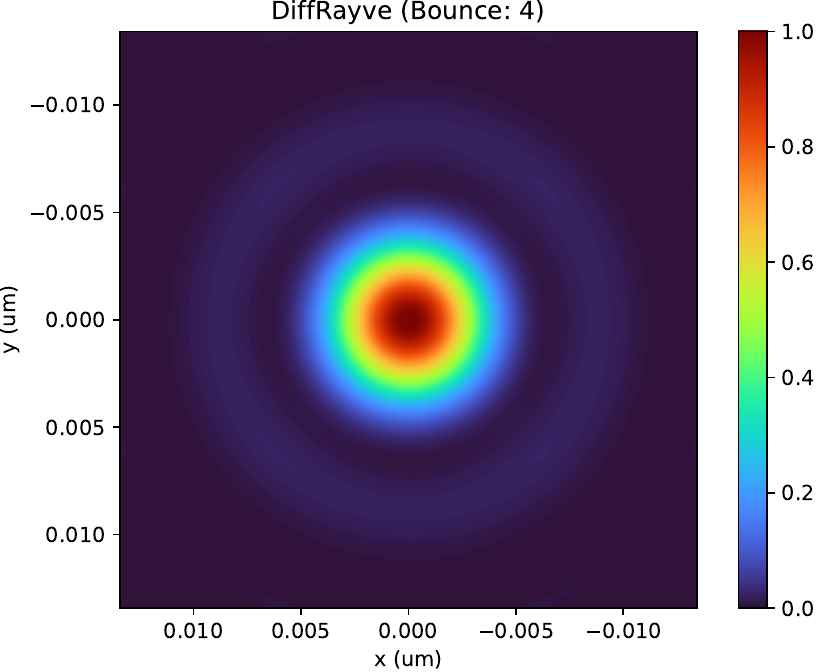}
    \caption{3 Bounces}
  \end{subfigure}
  \hfill
  \begin{subfigure}[b]{0.19\columnwidth}
    \centering
    \includegraphics[width=\linewidth]{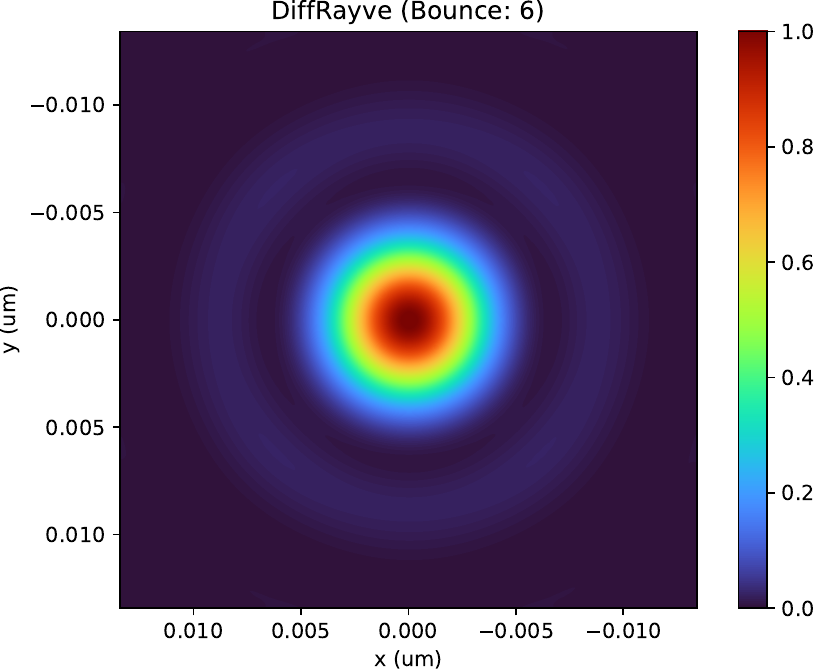}
    \caption{4 Bounces}
  \end{subfigure}
  \centering
  \begin{subfigure}[b]{0.19\columnwidth}
    \centering
    \includegraphics[width=\linewidth]{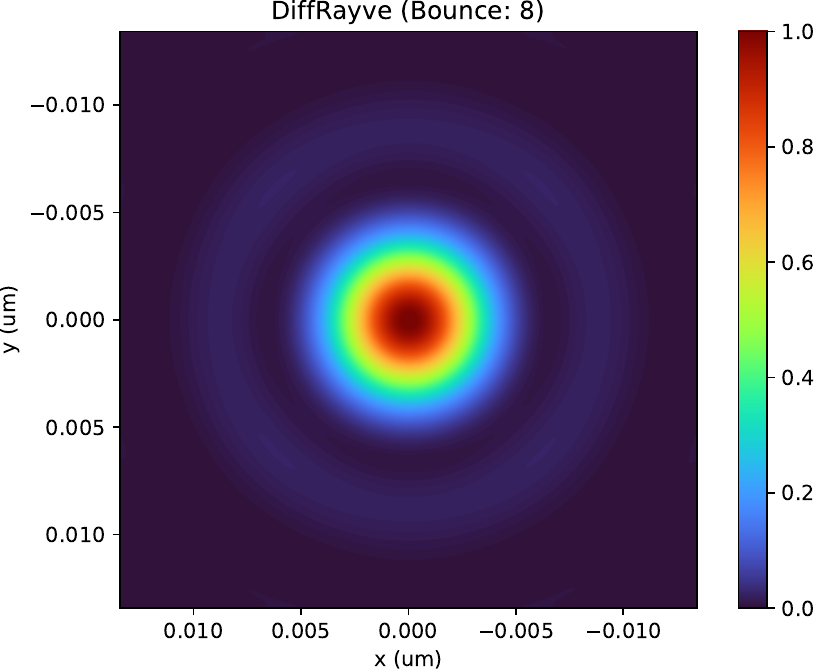}
    \caption{5 Bounces}
  \end{subfigure}
  \caption{\textbf{Max Ray Intersection Comparison.} We investigated the accuracy of different max ray intersections through the lens system and found that as long as rays are able to reach the last surface, the solution converges to the appropriate value. In our example, we normalize the PSFs; however, if field amplitudes are needed to optimize throughput, extended bounces may be necessary.}
  \label{fig:bounce-chart}
\end{figure}

Though our simulation can include long ray paths through the lens system,
we found that including additional bounces did not greatly influence the
result. In Figure \ref{fig:bounce-chart}, we plot the circular lens
response for paths of max length 1, 2, 4, 6, and 8. We found that as
long as the maximum length can make it to the final surface in the lens system, the field at the sensor is well characterized; however, if designers would like to compute light throughput and not normalize the point spread function (PSF), then more bounces may be necessary. This is out of scope of the present work but is supported by our simulation. We also report the runtime for each max bounce setting in Table \ref{tab:bounce-runtime}. Additional
bounces do increase the runtime of the solution, so if the normalized PSF is all that
is needed, it is best to set the value to 2 times the number of elements so that
all surfaces are intersected.

\subsection{Sampling Density}
\label{sec:sample-density}

\begin{figure}[h]
  \centering
  \begin{subfigure}[b]{0.19\columnwidth}
    \centering
    \includegraphics[width=\linewidth]{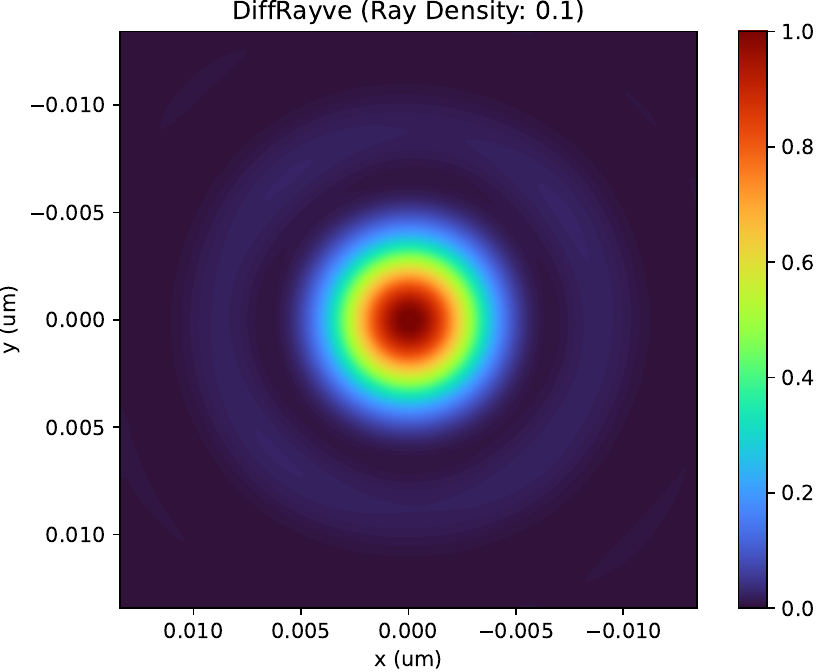}
    \caption{Density: 0.1}
  \end{subfigure}
  \hfill
  \begin{subfigure}[b]{0.19\columnwidth}
    \centering
    \includegraphics[width=\linewidth]{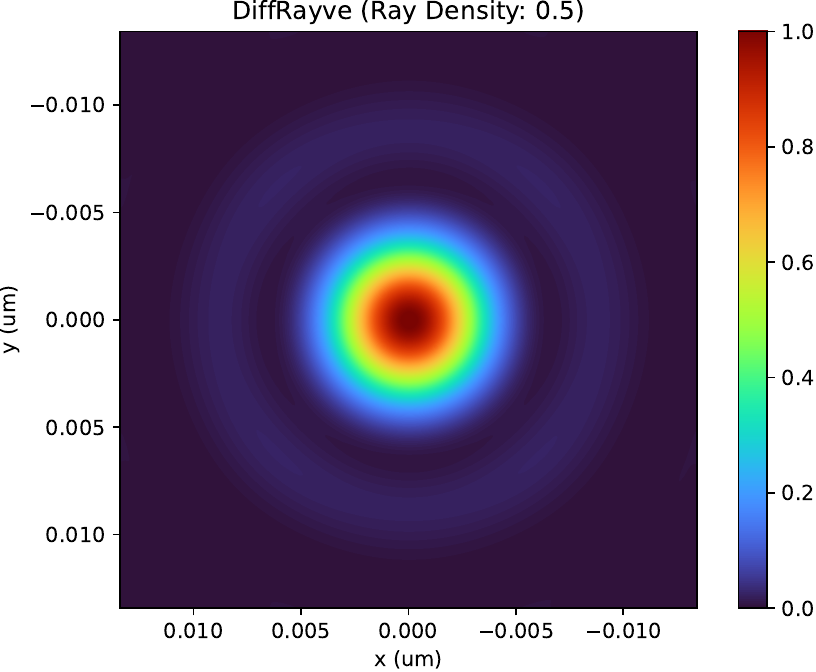}
    \caption{Density: 0.5}
  \end{subfigure}
  \centering
  \begin{subfigure}[b]{0.19\columnwidth}
    \centering
    \includegraphics[width=\linewidth]{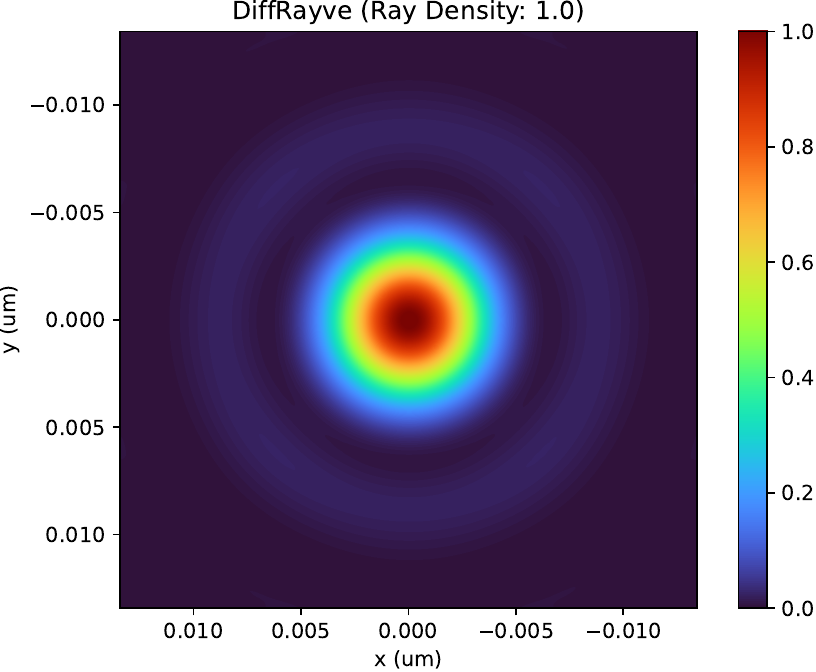}
    \caption{Density: 1.0}
  \end{subfigure}
  \hfill
  \begin{subfigure}[b]{0.19\columnwidth}
    \centering
    \includegraphics[width=\linewidth]{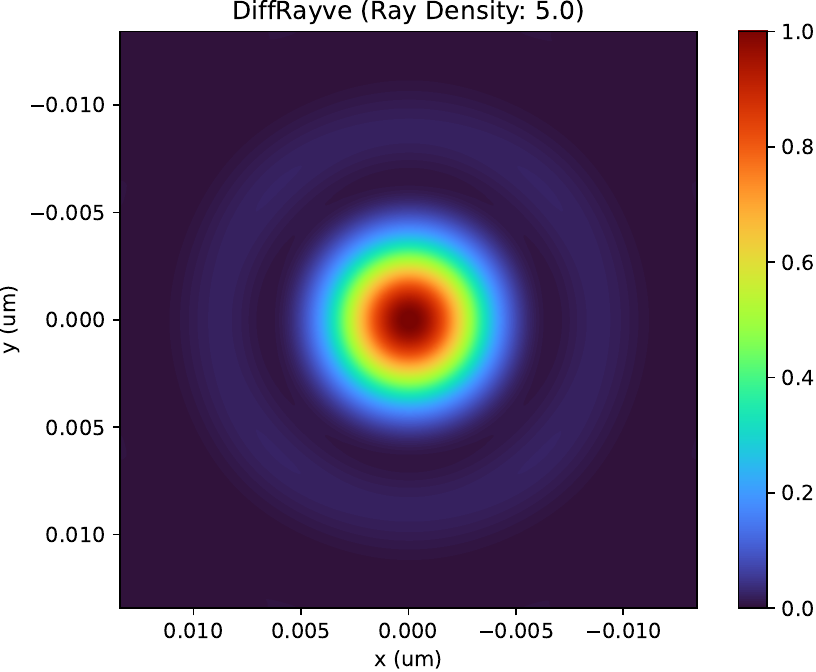}
    \caption{Density: 5.0}
  \end{subfigure}
  \centering
  \begin{subfigure}[b]{0.19\columnwidth}
    \centering
    \includegraphics[width=\linewidth]{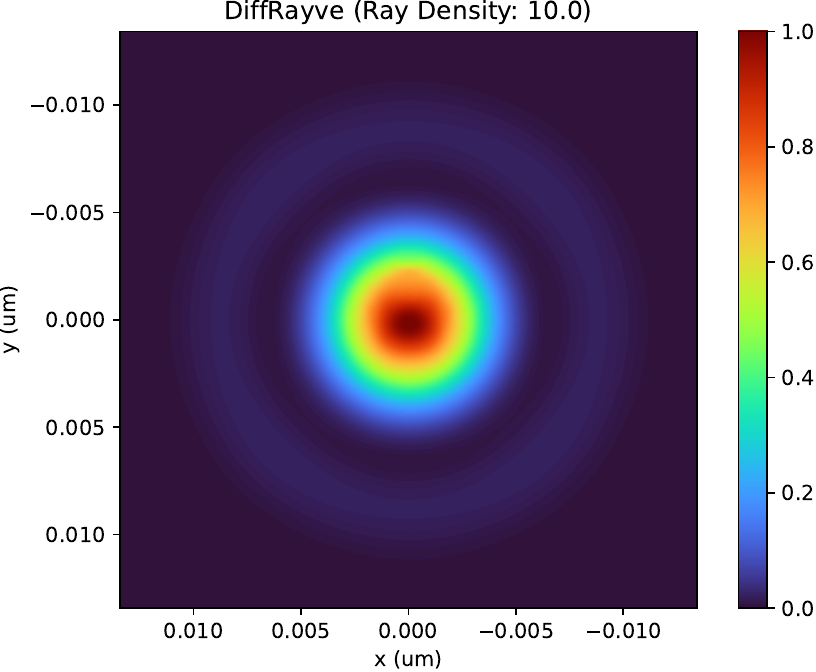}
    \caption{Density: 10.0}
  \end{subfigure}
  \caption{\textbf{Sampling Density Comparison.} We compare the sampling
  density of the incident plane wave against the accuracy of the calculated
  PSF. We find that few samples are needed in practice but maintain at 
  least 1 sample per wavelength in the main paper to ensure proper sampling
  on more complex systems without greatly increasing runtime.}
  \label{fig:density-chart}
\end{figure}

\begin{table}[h]
    \centering
    \caption{\textbf{Runtimes at Max Bounces.} We see that runtime grows with
    the number of intersections in the scene. This relationship is not quite linear
    due to the fact that the number of active rays decreases with each additional
    value as rays leave the lens system. We set the value to 2 times the number
    of optical elements in the paper because it was the best trade off between
    runtime and accuracy when computing normalized PSFs.}
    \label{tab:bounce-runtime}
    \begin{tabular}{c|ccccc}
        Bounces & 1 & 2 & 4 & 6 & 8 \\
        \hline
        Runtime (s) & 1.65 & 4.67 & 8.58 & 9.29 & 9.42 \\
    \end{tabular}
\end{table}

Similar to the bounce count, we explored different sampling densities
of the incident plane wave. The sampling density divides the wavelength by that
amount when figuring out the stratified sampling for the initial ray launch.
The density acts along a single dimension, so the area sampled is its square.
For example, 0.1 means that there is a sample for every 100 square wavelengths. 
Similarly, if the density is 5 then each square wavelength has 25 samples.
For smoothly varying surfaces with near normal incident directions, fewer samples
are necessary. This is difficult to know apriori, so we must pick a sufficiently
large sample. Figure \ref{fig:density-chart} shows that for the circular lens 
we are able to resolve the PSF with very few samples. In the main paper, we 
picked a middle ground value of 1 sample per wavelength. If we go higher, 
then runtime increases due to the quadratic increase in rays traced. Table
\ref{tab:density-runtime} shows that a density of 1 is a good balance as
runtime quickly increases.

\begin{table}[h]
    \centering
    \caption{\textbf{Runtime at Sampling Densities.} We list the runtime
    of the simulation for each ray density found in Figure \ref{fig:density-chart}.
    Due to the quadratic nature of this value, we find that runtime quickly increases.
    In the main paper, we pick a value of 1 sample per wavelength to maintain performance, while ensuring adequate sampling across different geometric surfaces.}
    \label{tab:density-runtime}
    \begin{tabular}{c|ccccc}
        Density & 0.1 & 0.5 & 1 & 5 & 10 \\
        \hline
        Runtime (s) & 0.595 & 1.61 & 4.57 & 107 & 432  \\
    \end{tabular}
\end{table}

\section{Multiple DOE Training Details}
\label{sec:mult-doe-details}

The PSF represents the response at the sensor plane for each direction incident on the aperture. We
locate the PSFs center using GO principles for the refractive
lens. Given incident angles $\theta$ and $\phi$ in spherical coordiante and the focal length $f$, the
PSFs location relative to the central axis is $\mathbf{d} = [f\tan(\theta)\cos(\phi), f\tan(\theta)\sin(\phi)]$. Each calculated PSF
has a width and height of $16\mu m$ and is computed on a $10 \times 10$ grid, giving a pixel pitch of $1.6 \mu m$.
The optimization goal is to shrink the PSF of three wavelengths of light
($350 nm$, $550 nm$, and $750 nm$) to reduce chromatic aberration. We use
SGD with a learning rate of $1e-3$, which is halved
after 7 updates. We stop optimization after 9 updates and found that a momentum of 0.9 gave the best results. 

In each gradient step, we compute the loss with respect to 16 incident angles (at $0^\circ$, $4^\circ$, and $8^\circ$ radially and at $45^\circ$ intervals rotationally) along with the 3 color channels and orthogonal polarizations (Equation~6) to 
computed the unpolarized PSF. This setup results in 96 SBR forward simulations and backpropagation each step. We sample at 1 ray per wavelength
so each simulation, on average, propagates 4 million GO rays through the system. 

To compute the loss function, we first normalize the PSF, giving $\hat{\mathbf{P}}_k$ for a given color channel. Each pixel, $p^{(k)}_{i, j}$, on the 2D grid indexed by $i$ and $j$, is then assigned a normalized coordinate, $x^\prime_{i, j}$ and $y^\prime_{i, j}$, on a square
from $-1$ to $1$. The pixels intensity is them weighted by
its radial distance $\rho_{i, j} = (x^\prime_{i, j})^2 + (y^\prime_{i, j})^2$. This weighting encourages the PSF to reduce in size towards the center. We also include a loss term by computing the variance $\sigma^2_k = \frac{1}{100}\sum_{i=1}^{10}\sum_{j=1}^{10}(\rho_{i,j}\hat{p}^{(k)}_{i,j} - \mu_k)^2$, where $\mu_k$ is the weighted mean and $\hat{p}_{i, j}$ is the normalized pixel intensity. This additional term further reduces the spread of the PSF. In total, the loss at each gradient step is 
\begin{equation}
    L = \frac{1}{16}\sum_{k=1}^{16\cdot3}\Big[\sigma^2_k+\frac{1}{100}\sum_{i=1}^{10}\sum_{j=1}^{10}\rho_{i,j}\hat{p}^{(k)}_{i,j}\Big].
    \label{eq:mult-doe-loss-supplement}
\end{equation}

\section{Out of Distribution Angles}
\label{sec:ood-angles}

\begin{figure}[h]
  \centering
  \begin{subfigure}[b]{0.19\columnwidth}
    \centering
    \includegraphics[width=\linewidth]{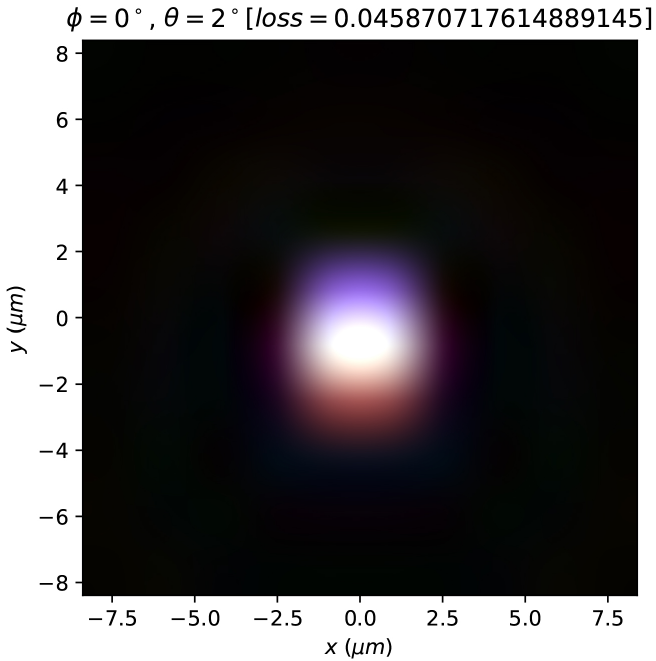}
  \end{subfigure}
  \hfill
  \begin{subfigure}[b]{0.19\columnwidth}
    \centering
    \includegraphics[width=\linewidth]{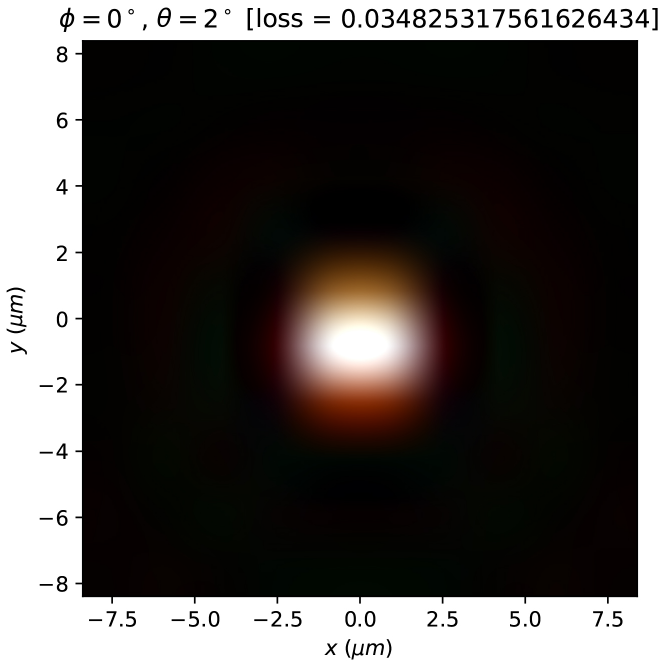}
  \end{subfigure}
  \centering
  \begin{subfigure}[b]{0.19\columnwidth}
    \centering
    \includegraphics[width=\linewidth]{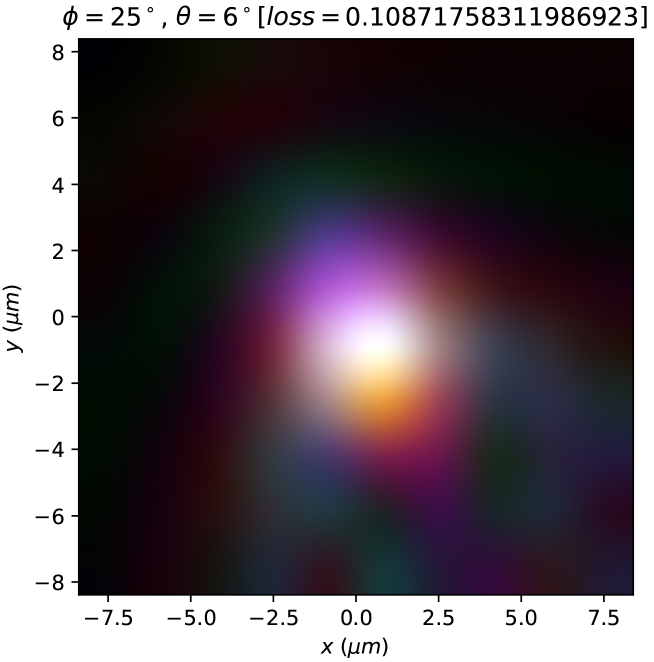}
  \end{subfigure}
  \hfill
  \begin{subfigure}[b]{0.19\columnwidth}
    \centering
    \includegraphics[width=\linewidth]{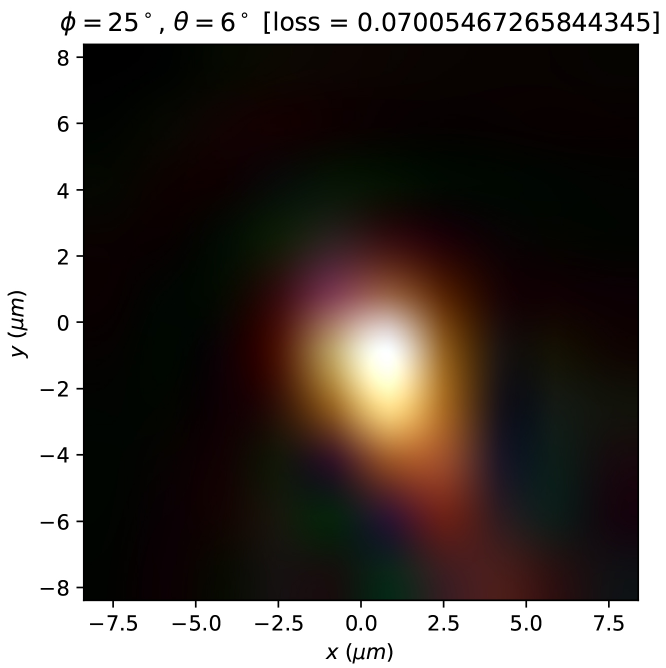}
  \end{subfigure}
  \centering
  \begin{subfigure}[b]{0.19\columnwidth}
    \centering
    \includegraphics[width=\linewidth]{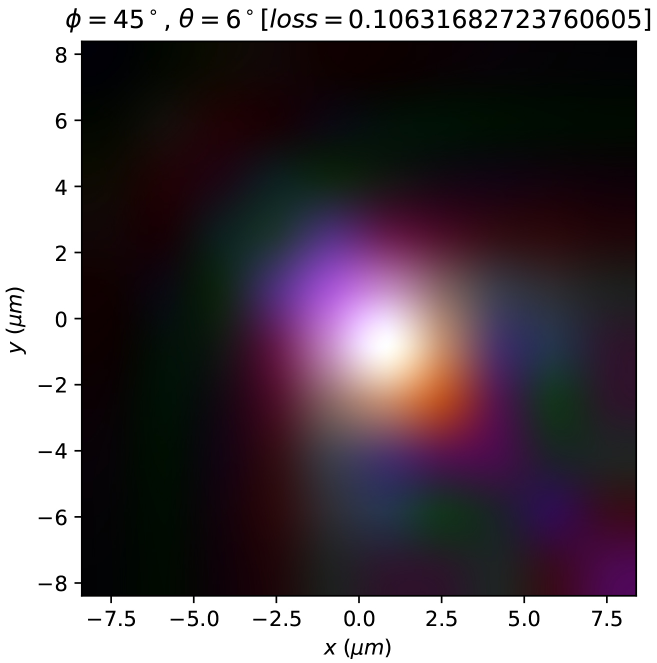}
  \end{subfigure}
  \hfill
  \begin{subfigure}[b]{0.19\columnwidth}
    \centering
    \includegraphics[width=\linewidth]{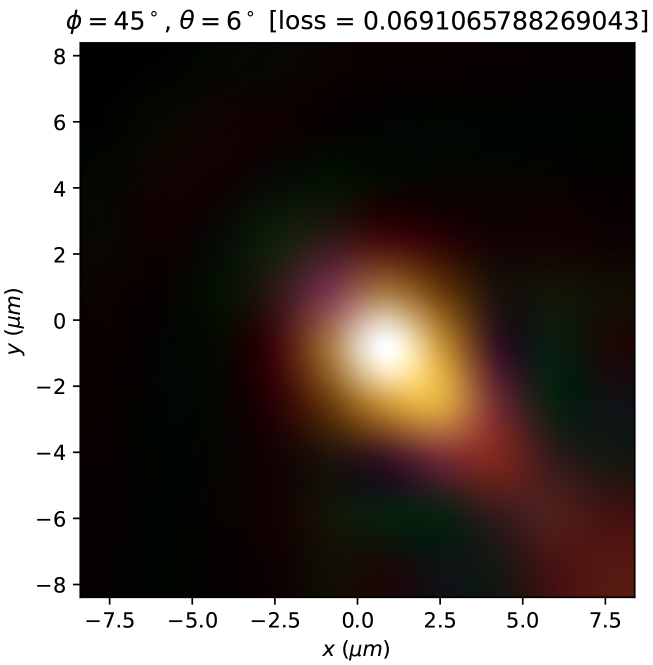}
  \end{subfigure}
  \centering
  \begin{subfigure}[b]{0.19\columnwidth}
    \centering
    \includegraphics[width=\linewidth]{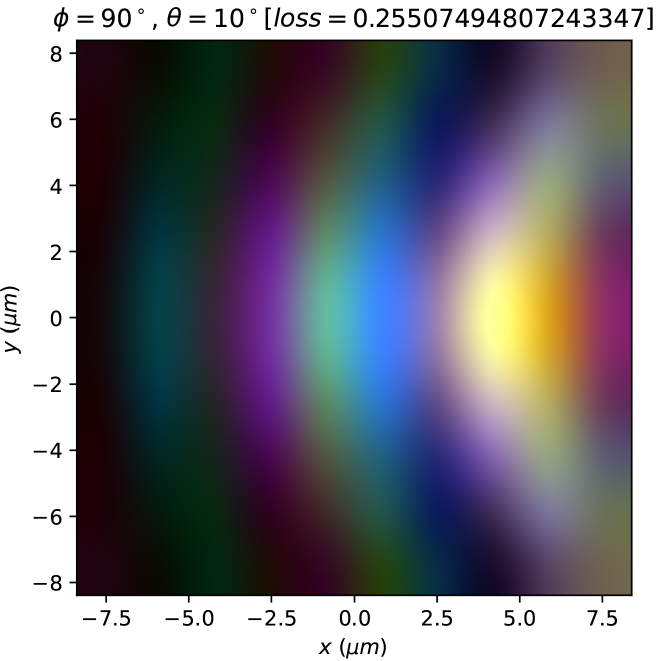}
  \end{subfigure}
  \hfill
  \begin{subfigure}[b]{0.19\columnwidth}
    \centering
    \includegraphics[width=\linewidth]{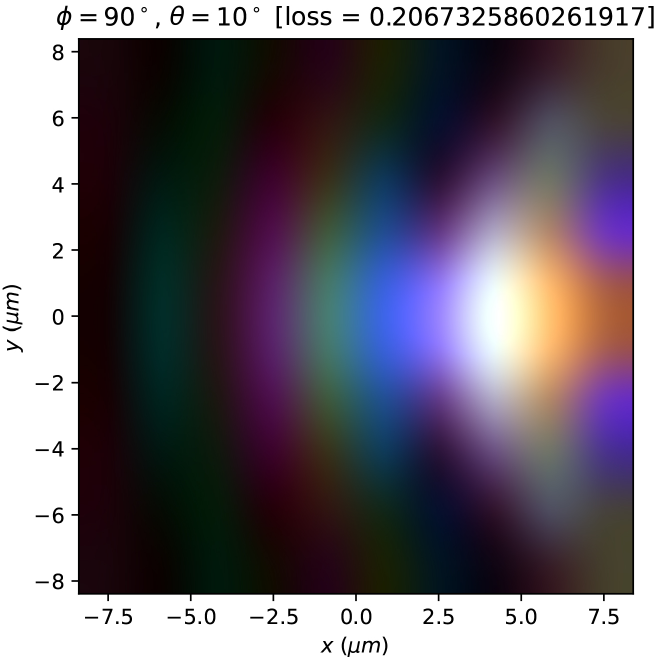}
  \end{subfigure}
  \centering
  \begin{subfigure}[b]{0.19\columnwidth}
    \centering
    \includegraphics[width=\linewidth]{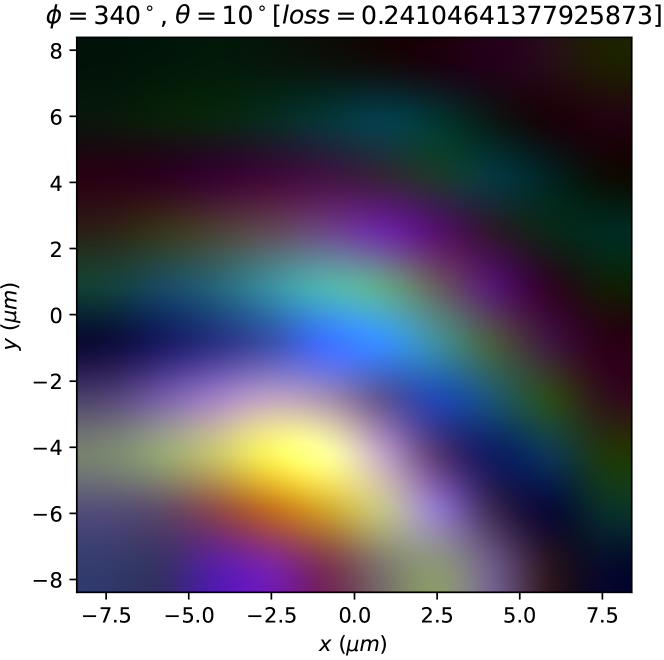}
  \end{subfigure}
  \hfill
  \begin{subfigure}[b]{0.19\columnwidth}
    \centering
    \includegraphics[width=\linewidth]{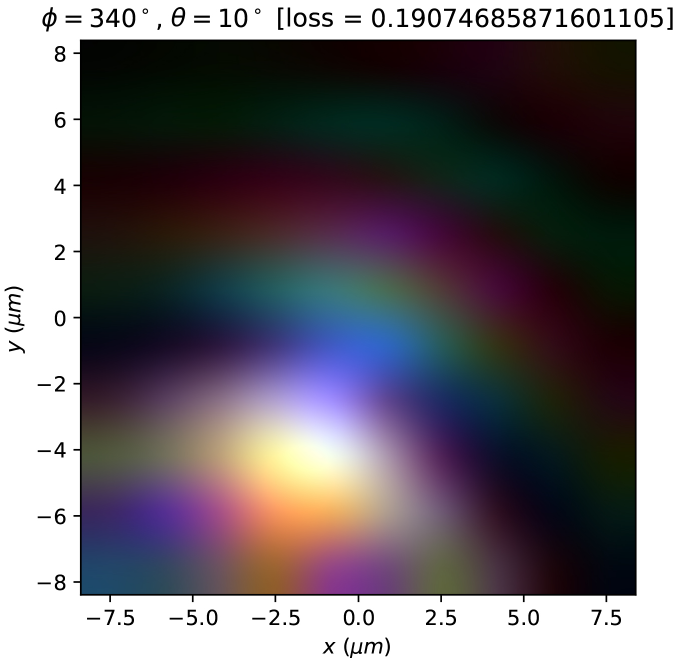}
  \end{subfigure}
  \caption{\textbf{Out of Distribution PSFs.} We plot the before (left) and 
  after (right) PSFs of angle not included in the optimization step of
  the multiple DOE system in the main paper results. The losses, as defined
  in the main paper, are reported for each PSF. We demonstrate an improvement
  even when the angle was not optimized for.}
  \label{fig:out-of-distribution}
\end{figure}

In the main paper, we demonstrated that our differentiable simulation
was able to optimize multiple diffractive optical elements in a lens system. To optimize wide off axis field of view angles, we picked 
fixed angles. This is not a limitation of our method and any incident angle
can be considered. Though the goal of the experiment was not to propose a 
lens system and rather demonstrate the adjoint simulation, we investigate the lens system's performance at non optimized angles. Due to symmetry, we only select a few angles
around the field of view and compute their PSFs (Figure \ref{fig:out-of-distribution}). Losses as defined in the main paper are reported for each PSF and we plot values before and after optimization. We find that even though these angles were not directly optimized,
performance is improved and the PSFs are condensed. For angles within the original
field of view, performance is the strongest as there is better coverage in the optimization. For wider angles, there is a slight improvement; however, the improvement is worse than other test angles due to lack of geometric overlap in the training set.
This result demonstrates the usefulness of our method as new diffractive systems can be designed efficiently without the needing to compute PSFs at all angles in the field of view.

\section{Baseline Details}
\label{sec:baseline-details}

In this section, we discuss the details of how we ran other
simulation baselines: Chromatix \cite{deb_chromatix_2023, deb_chromatix_2025} and DeepLens \cite{yang_curriculum_2024, yang_end--end_2024}. In some instances, small modifications needed to be
made to the code base to make sure inputs were equivalent. 

\subsection{Fraunhofer Diffraction Equations}

The circular lens produces an
Airy disk \cite{goodman_introduction_2017} calculated by
\begin{equation}
  E(\theta) \propto \Big|\frac{2J_1(\theta)}{\theta}\Big|^2;\;
  \theta = \frac{\pi d}{\lambda} \frac{\rho}{f}
  \label{eq:airy-disk}
\end{equation}
where $d$ is the lens diameter, $f$ is the focal length, and $J_1$ is
the first order Bessel function of the first kind. We compute a $1mm$
diameter plano-convex lens, with a $10 mm$ focal length, index of refractions of 1.5, and incident wavelength of $550nm$. The sinusoidal
phase grating also has a known solution for Fraunhofer diffraction given
by
\begin{multline}
  E(x, y, z) \propto \sum_{q=-\infty}^{\infty}
  J_q^2(\frac{m}{2}) \cdot \mathrm{sinc}^2(\frac{2w}{\lambda
  z}(x - qf_0\lambda z))\\
  \mathrm{sinc}^2(\frac{2wy}{\lambda z}),
  \label{eq:sinusoidal-grating}
\end{multline}
where $w$ is the element half width, $f_0$ is grating frequency, $m$ is
the mean phase difference across the lens
thickness, and $J_q$ is the $q$ order Bessel function
of the first kind. Our experiments set $w$ to $5 \mu m$, $f_0$ to
$0.5 \mu m^{-1}$, and $m$ to $\frac{5 \pi \sqrt{2}}{5.5e-1}$.
Again, the incident wavelength is $550 nm$. We adapt the grating slightly to account
for the cosine decay for large sensors along with the smearing produced by
breaking the paraxial approximation. The new equation reads as
\begin{multline}
  E(x, y, z) \propto \cos(\theta)^4 \sum_{q=-\infty}^{\infty}
  J_q^2(\frac{m}{2}) \cdot \mathrm{sinc}^2(\frac{2w}{\lambda
  z} \\
  \cos(\theta)(x - qf_0\lambda z))\mathrm{sinc}^2(\frac{2wy}{\lambda z}),
  \label{eq:sinusoidal-grating-corrected}
\end{multline}
where $\theta$ is the angle between the central axis and the sensor location.

Now consider the polarized scattered fields produced by a circular plano-convex lens. For a field with initial polarization
along the x axis, the analytical solution the field at the focal length is
\begin{equation}
  E_x(r, \phi) \propto I_0(\rho') + I_2(\rho')\cos(2\phi)
  \label{eq:vector-x},
\end{equation}
\begin{equation}
  E_y(r, \phi) \propto I_2(\rho')\sin(2\phi)
  \label{eq:vector-y},
\end{equation}
and
\begin{equation}
  E_z(r, \phi) \propto I_1(\rho')\cos(\phi),
  \label{eq:vector-z}
\end{equation}
where $\rho'$ is the scaled distance $k_0 \sqrt{x^2+y^2}$ and $\phi$ is the
angle given by $\tan^{-1}(\frac{y}{x})$ with $z$ along the central lens axis.
The numerical aperture is $NA$ and the index of refraction is $n$. The three $I$
values are then given by the integrals
\begin{equation}
  I_0(\rho^\prime) = \int_0^\alpha \sqrt{\cos(\theta)}\sin(\theta)(1+\cos(\theta))J_0(\rho^\prime\sin(\theta))d\theta
  \label{eq:int-0},
\end{equation}
\begin{equation}
  I_1(\rho^\prime) = \int_0^\alpha \sqrt{\cos(\theta)}\sin^2(\theta)J_1(\rho^\prime\sin(\theta))d\theta
  \label{eq:int-1},
\end{equation}
and
\begin{equation}
  I_2(\rho^\prime) = \int_0^\alpha \sqrt{\cos(\theta)}\sin(\theta)(1-\cos(\theta))J_2(\rho^\prime\sin(\theta))d\theta,
  \label{eq:int-2}
\end{equation}
with $\alpha = \tan^{-1}(\frac{d}{2f})$.

\subsection{Chromatix}

\subsubsection{Circular Lens}
In this experiment, we use a $1mm$ diameter ($d$) plano-convex lens, with a $10mm$ focal length ($f$), and index of refraction ($n$) of 1.5, same as Diffrayve. The sensor size ($\approx 27\mu m$) and sensor resolution ($62 \times 64$) are the same as diffrayve. Chromatix being a wave optics framework, we initialize the wavefront as a square electromagnetic vector field of wavelength $550nm$ of size slightly larger than the diameter of the lens (i.e. $1.1d$) and resolution $128 \times 128$, as opposed to a grid of rays in Diffrayve. This wavefront is initially polarized along x-axis. A circular pupil limits the wavefront to the bounds of the lens. Finally, the polarized field from the lens itself is computed using a high numerical aperture lens, whose numerical aperture (NA) is computed as $n sin(\alpha)$, where $\alpha$ is half-angle of the maximum cone of light that can enter the lens, i.e. $\frac{d}{2f}$. Figure \ref{fig:chromatix_circular} contains the code. 

\begin{figure}[H]
\begin{minted}[fontsize=\tiny, breaklines]{python}
    import chromatix as cx

    def chromatix_circular(diameter, extent, focal_length, res, wavelength):
        _res = 128
        _dx = diameter * 1e6 * 1.1 / _res
        _n = 1.5
        _NA = _n * diameter / (focal_length * 2)
        cx_field = cx.plane_wave(
            shape=(_res, _res), 
            dx=_dx, 
            spectrum=wavelength * 1e6, scalar=False, amplitude=[0, 0, -1.0]
        )
        cx_field = cx.circular_pupil(cx_field, w=diameter * 1e6)
        cx_field = cx.high_na_ff_lens(
            cx_field,
            f=focal_length * 1e6,
            n=_n,
            NA=_NA,
            output_shape=(res, res, 3),
            output_dx=extent / res * 1e6,
        )
        cx_field = np.abs(cx_field.u.squeeze())
        return cx_field
\end{minted}
\caption{Chromatix Code for Circular Lens}
\label{fig:chromatix_circular}
\end{figure}

\subsubsection{Sinusoidal Grating}

\begin{figure}[H]
\begin{minted}[fontsize=\tiny, breaklines]{python}
    import chromatix as cx

    def chromatix_sinusoid(width, buffer, wavelength, n, f0, d, z):
        _res = 8400
        _dx = 2 * width * buffer * 1e6 / _res
        cx_field = cx.plane_wave(
            shape=(_res, _res),
            dx=_dx,
            spectrum=wavelength * 1e6
        )
        cx_field = cx.square_pupil(cx_field, w=2 * width * 1e6)
        cx_field = cx.sinusoid_grating(
            cx_field, n_grating=n, period=f0 / 1e6, thickness=d * 1e6
        )
        cx_field = cx.transfer_propagate(cx_field, z=z * 1e6, n=1, N_pad=0)
        return cx_field
\end{minted}
\caption{Chromatix Code for Sinusoid Grating}
\label{fig:chromatix_sinusoid}
\end{figure}
In this experiment, we model a sine grating of half-width $w = 5\mu m$, period $f_0 = 0.5 \mu m^{-1}$ and thickness $d = 5\sqrt{2} \mu m$ as a linear operator, as opposed to Diffrayve, where the grating is realized as a glass slab lens with a heightmap texture on the front face to modulate the equivalent phase change. However, the incident wavefront (and therefore the sensor) are larger than the width of the grating (by a factor of 30, stored in the variable \textit{buffer}), to capture a significant size of the output pattern. The incident wavefront has a resolution of $8400 \times 8400$, and it is non-zero only in the region limited by the grating. After the phase change, the wavefront is propagated to the sensor at depth $z = 100 \mu m$, using convolutional Fresnel propagation (CV-FR). Figure \ref{fig:chromatix_sinusoid} contains the code. 

\subsection{DeepLens}

\subsubsection{Circular Lens}

\begin{figure}[H]
\begin{minted}[fontsize=\tiny, breaklines]{python}
    import torch
    import numpy as np
    from deeplens.hybridlens import HybridLens
    from deeplens.optics.monte_carlo import forward_integral
    import torch.nn.functional as F
    from deeplens.optics.wave import AngularSpectrumMethod

    #Hybrid Lens
    hlens = HybridLens(
        filename="./datasets/lenses/custom/circular_hybrid.json",
        dtype=torch.float64
    )

    #Initial Ray Grid
    ray_res = 20
    ks = hlens.geolens.sensor_res[0]
    hrays = hlens.geolens.sample_grid_rays(
        num_grid=ray_res,
        num_rays=1,
        wvln=0.55
    )
    hrays.coherent = True

    #Optimal Position of Identity DOE
    hlens.doe.d = 950

    # Step 1: Ray Trace to DOE and calculate integral on it
    hrays, _ = hlens.geolens.trace(hrays)
    hrays = hrays.prop_to(hlens.doe.d)
    hrays = hrays.reshape(-1,1) # Convert [N, N, spp] to [N*N, spp]
    hfield = forward_integral(hrays,
        ps=hlens.doe.fab_ps,
        ks=hlens.doe.res[0],
        coherent=True
    )
    
    # Step 2: Apply phase map (identity) to computed field (wavefront)
    _wavefront = torch.sum(hfield,axis=0)
    phase_map = torch.flip(hlens.doe.get_phase_map(0.55), [-1, -2])
    _wavefront = _wavefront * torch.exp(1j * phase_map)
    
    # Step 3: Propagate wavefront to sensor
    h, w = _wavefront.shape
    sensor_field = AngularSpectrumMethod(_wavefront,
        z=hlens.geolens.d_sensor - hlens.doe.d,
        wvln=0.55,ps=hlens.doe.fab_ps,padding=True
    )
    sensor_field = sensor_field.abs() ** 2
    _result = sensor_field.squeeze().cpu().numpy()
    _result = (_result - _result.min()) / (_result.max() - _result.min())
\end{minted}
\caption{DeepLens Code for Circular Lens}
\label{fig:deeplens_circular}
\end{figure}

For this experiment, we construct a plano-convex lens of diameter $d = 50mm$, radius of curvature $r = 500mm$ and refractive index $n = 1.5$. A sensor of resolution $64 \times 64$ and radius $1.35mm$ is placed at the focal length $1m$. Note that these are the same parameters as Diffrayve and Chromatix, only scaled up by a factor of 100, since DeepLens uses the unit $mm$ by default in its operators, except wavelength, which is stored in $\mu m$, and we set it equal to $0.55 \mu m$. 

DeepLens, primarily being a geometric optics (GO) framework, does not yield the characteristic PSF pattern on the sensor plane, as all rays trace to a single point at the center. So, we instead use a hybrid lens (combination of plano-convex lens and a diffractive optical element (DOE) with identity phase transformation). The DOE has resolution $64 \times 64$. The corresponding input file is detailed in Figure \ref{fig:deeplens_lens_circular}. In this hybrid configuration, the initial ray grid of resolution $20 \times 20$ and spp 1, is traced till the DOE, at which point it is transformed into a complex wavefront using Monte-Carlo integration, which is finally propagated till the sensor plane using angular spectrum method (ASM). Figure \ref{fig:deeplens_circular} contains the code. 

In this experiment, the normalized magnitude of the wavefront at the sensor depends on the size and location/depth of the DOE. We find that a DOE of size $1.92mm$ at a depth of $950mm$ yields the best results.

\begin{figure}[H]
\begin{minted}[fontsize=\tiny, breaklines]{json}
                {
                    "info": "DeepLens Circular",
                    "r_sensor": 1.35,
                    "sensor_res": [64, 64],
                    "surfaces": [
                        {
                            "idx": 1,
                            "type": "Spheric",
                            "r": 50,
                            "roc": 500,
                            "mat1": "air",
                            "mat2": "1.5/50",
                            "d_next": 2.5
                        },
                        {
                            "idx": 2,
                            "type": "Plane",
                            "r": 50,
                            "mat1": "1.5/50",
                            "mat2": "air",
                            "d_next": 1000.0
                        }
                    ],
                    "DOE": {
                        "type": "DOE",
                        "d": 2.5,
                        "res": [
                            64,
                            64
                        ],
                        "fab_ps": 0.03,
                        "param_model": "pixel2d",
                        "wvln0": 0.55
                    }
                }
\end{minted}
\caption{Lens input \textit{circular\_hybrid.json} for DeepLens}
\label{fig:deeplens_lens_circular}
\end{figure}

\subsubsection{Sinusoidal Grating}
DeepLens already contains implementations of a few types of diffractive surfaces. We created a similar diffractive surface with sinusoid phase transformation, $\phi = \frac{m}{2} sin(2\pi f_0 x)$ at depth $z = 0$. The grating half-width $w$ and period $f_0$ are same as Chromatix and Diffrayve, and $m = \frac{5 \pi \sqrt(2)}{5.5e-1}$. The resolution of the DOE is $160 \times 160$. Similar to Chromatix, the size of the incident wavefront and the sensor is 30 times the size of DOE, and thus a resolution of $4800 \times 4800$, to capture the entire output which extends beyond the limits of the DOE. Contrary to the circular lens experiment, we use Fresnel Propagation in this experiment to propagate the phase modified wavefront to the sensor plane at $z = 0.1mm$. Code for this part is in Figure \ref{fig:deeplens_sinusoid}.  

\begin{figure}[H]
\begin{minted}[fontsize=\tiny, breaklines]{python}
    import torch
    import numpy as np
    from deeplens.optics.diffractive_surface.sinusoid import Sinusoid
    from deeplens.optics.wave import ComplexWave

    # Parameters
    width = 1e-2 # mm
    buffer = 30
    f = 500 # mm^{-1}
    wvln = 0.55 # um
    d = 5 * np.sqrt(2) * 1e-3 # mm
    n = 1.5
    waveSamples = 4800
    samples = waveSamples // buffer
    dx = width / samples
    m = 2 * np.pi * (n-1) * d / wvln * 1e3
    # Final Propagation Depth
    z = 0.1

    # Sine Grating
    grating = Sinusoid(d=0.0, f=f, m=m, res=(samples,samples), fab_ps=dx, wvln0=wvln)
    wavefront = ComplexWave.plane_wave_square(wvln=wvln, phy_size=(width*buffer,width*buffer), res=(waveSamples,waveSamples), pupil_size=(width,width))
    wavefront = grating.forward(wavefront)
    wavefront = wavefront.prop_to(z=z)

    _result = wavefront.u.squeeze().cpu().numpy()
    _result = np.abs(_result) ** 2
    _result = _result / np.max(_result)
\end{minted}
\caption{DeepLens Code for Sinusoid Grating}
\label{fig:deeplens_sinusoid}
\end{figure}

\section{Vector Diffractive Optical Element Optimization}
\label{sec:add-vector-opt-result}

\begin{figure}[ht]
    \centering
    \begin{subfigure}{0.49\linewidth}
        \includegraphics[width=1.0\linewidth]{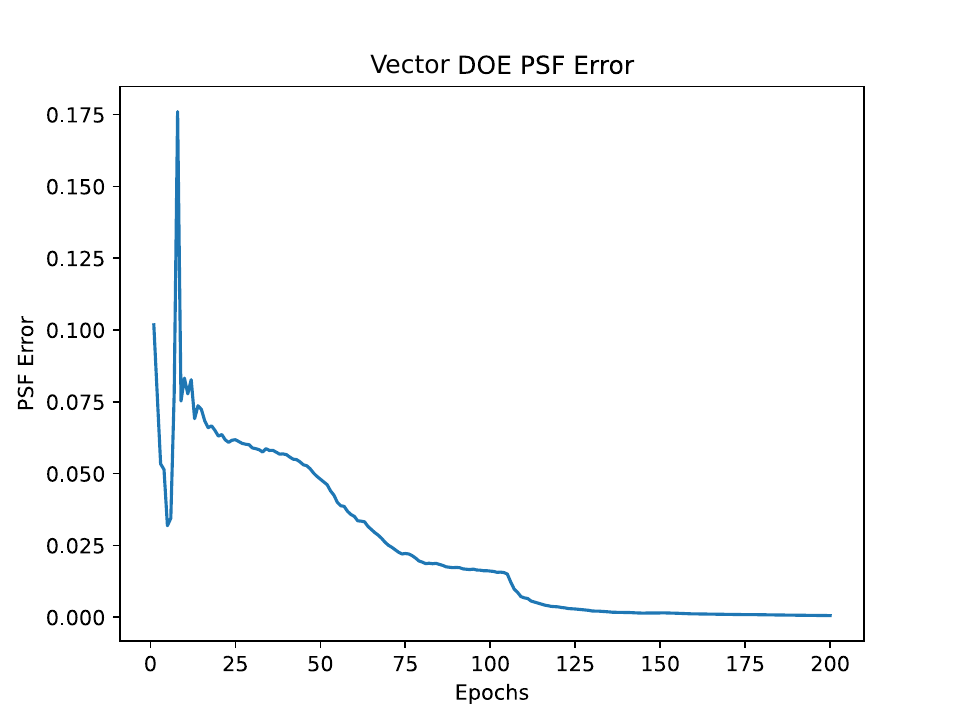}
    \end{subfigure}
    \begin{subfigure}{0.49\linewidth}
        \includegraphics[width=1.0\linewidth]{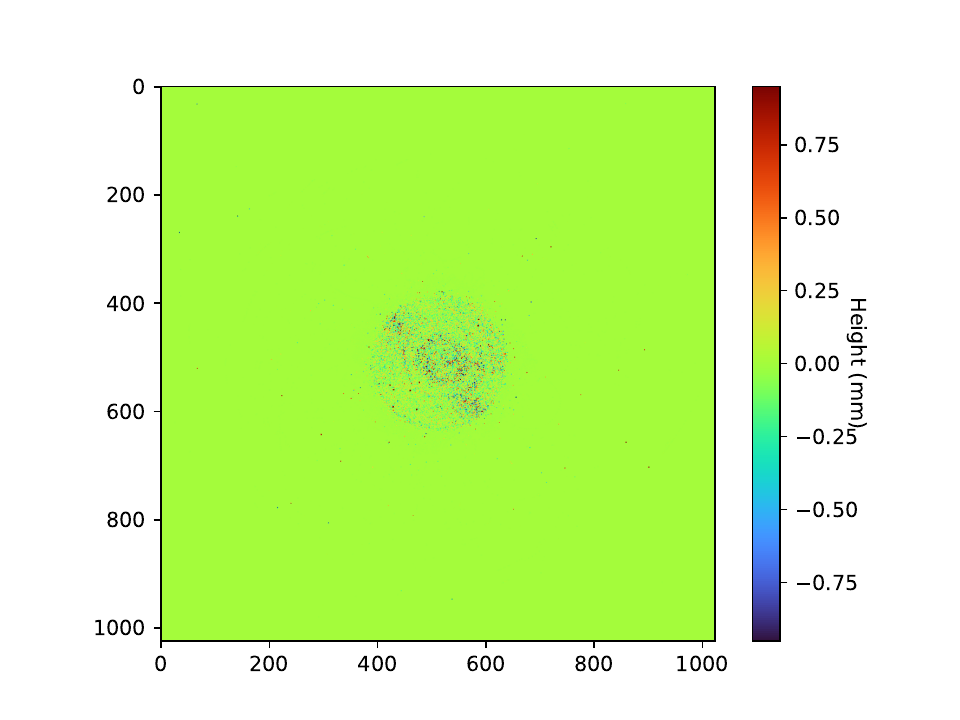}
    \end{subfigure}
    \caption{\textbf{Vector Optimization Loss and DOE.} We plot the loss curve for
    the vector field optimization experiment. This figure demonstrates
    that the gradients from our simulation are able to efficiently decrease
    the loss from 0.1 to 0.0005. The DOE phase mask is then plotted as the per pixel
    phase offset to transform the y component of the electric field to match the x component.}
    \label{fig:vec-loss}
\end{figure}

\begin{figure}[h]
  \centering
  \begin{subfigure}[b]{0.32\columnwidth}
    \centering
    \includegraphics[width=\linewidth]{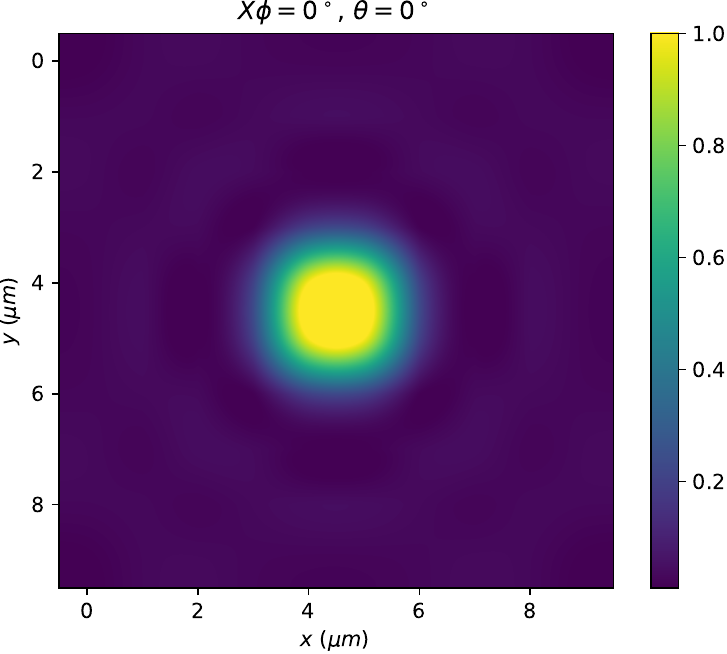}
  \end{subfigure}
  \hfill
  \begin{subfigure}[b]{0.32\columnwidth}
    \centering
    \includegraphics[width=\linewidth]{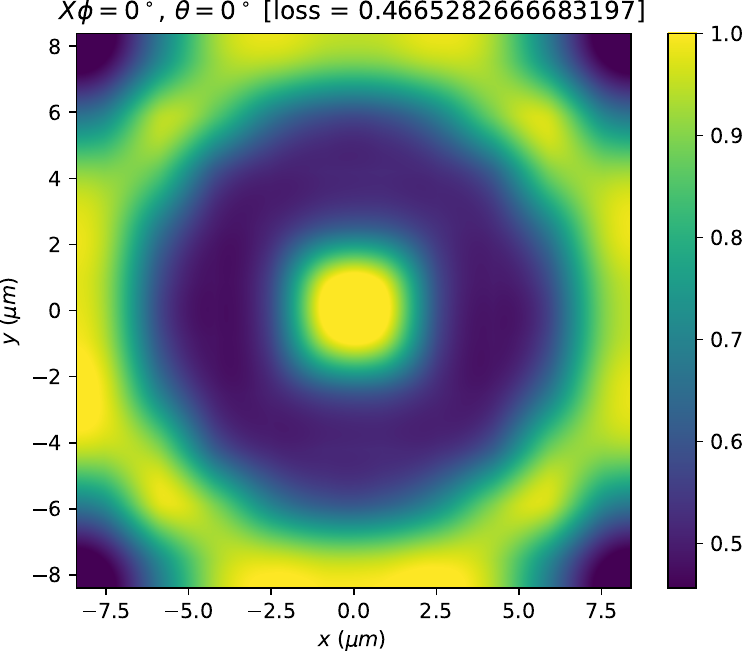}
  \end{subfigure}
  \centering
  \begin{subfigure}[b]{0.32\columnwidth}
    \centering
    \includegraphics[width=\linewidth]{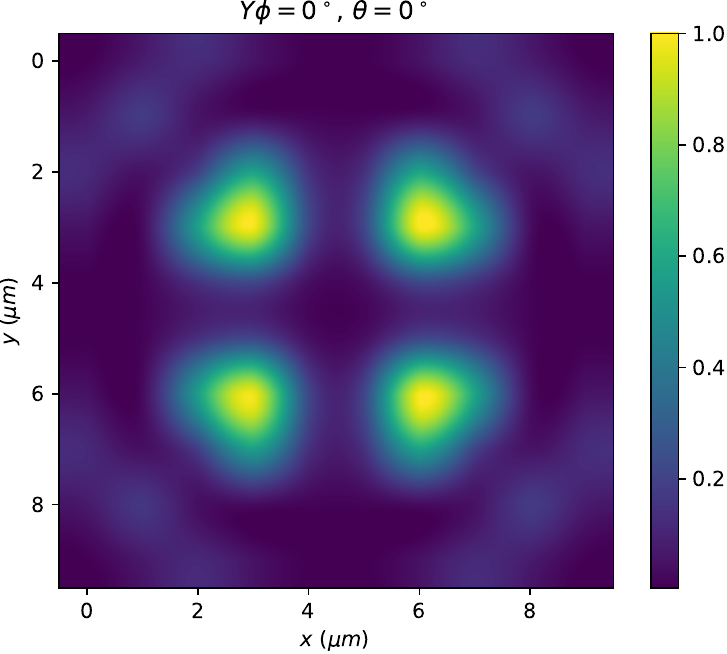}
  \end{subfigure}
  \hfill
  \begin{subfigure}[b]{0.32\columnwidth}
    \centering
    \includegraphics[width=\linewidth]{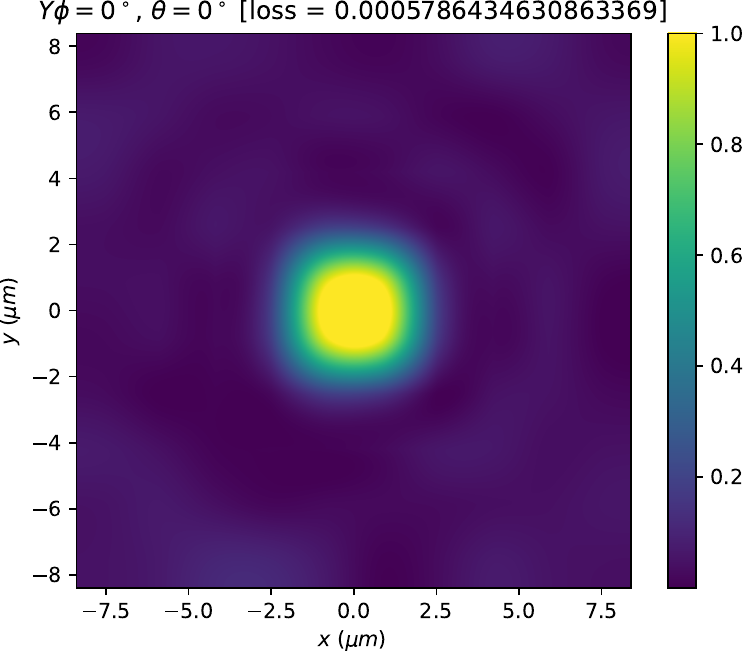}
  \end{subfigure}
  \centering
  \begin{subfigure}[b]{0.32\columnwidth}
    \centering
    \includegraphics[width=\linewidth]{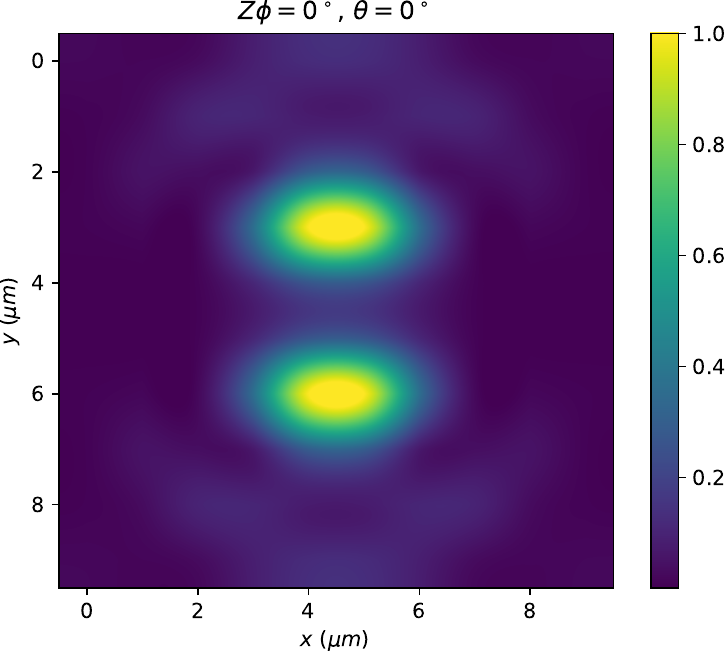}
  \end{subfigure}
  \hfill
  \begin{subfigure}[b]{0.32\columnwidth}
    \centering
    \includegraphics[width=\linewidth]{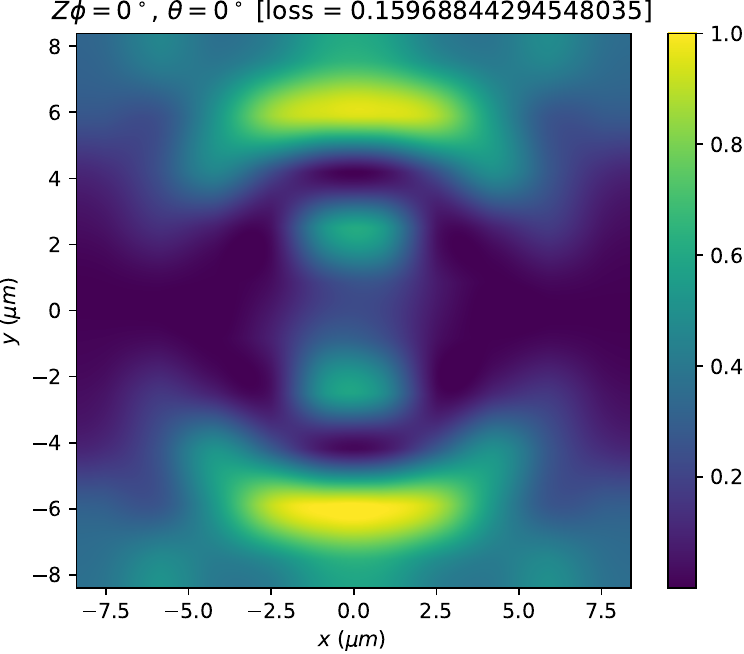}
  \end{subfigure}
  \caption{\textbf{Vector PSF Optimization.} Our method is able to calculate
  the full 3D vector field at the sensor along with its gradients. We were
  able to optimize a DOE so that the y component of the field matches the x
  component produced by a circular plano-convex lens. We see that the resulting
  PSF matches closely with the desired value demonstrating the capability of
  our method.}
  \label{fig:vec-psfs}
\end{figure}
Our method is able to compute either unpolarized or polarized light
in an optical system. In this section, we demonstrate that our simulation
can model diffractive elements that affect the polarized field. We consider a
system with a circular plano-convex lens and a pixel mask diffractive element.
The geometry matches that in the achromatic multiple DOE result in the main
paper with the binary DOE removed. Only a single wavelength of light at $550 nm$
was simulated.
The incident field is linearly polarized along the x axis and only paraxial
angles are considered. We optimized the pixel mask for 200 gradient steps
using stochastic gradient descent \cite{kiefer_stochastic_1952} with
a learning rate of $1e{-1}$ and momentum set at $0.9$. Optimization runtime was around 1 minute and used less than 1 GB of memory. The loss curve is plotted in Figure \ref{fig:vec-loss}. The final mask is shown in Figure \ref{fig:vec-loss}
and the resulting normalized PSFs along the $x$, $y$, and $z$ axes are plotted in 
Figure \ref{fig:vec-psfs}. The loss function was the mean squared error between the normalized y
and the x component of the field before the system was optimized. The optimization was
able to decrease the loss from 0.1 to 0.0005 forcing the y-axis field to match that x-axis; however, due to the lack of constraints in the loss function, the x and z components were affected. This experiment demonstrates the flexibility of our method
to compute gradients for optimization of not only unpolarized but also polarized fields.


\end{document}